\documentclass[preprint,aps,floats,nofootinbib,amssymb]{revtex4}
\usepackage{epsf,epsfig}

\newcommand{\be}{\begin{equation}}
\newcommand{\ee}{\end{equation}}
\newcommand{\bea}{\begin{eqnarray}}
\newcommand{\eea}{\end{eqnarray}}

\renewcommand{\check}{({\bf This statement needs to be checked!!})}

\newcommand{\lsim}{
\mathrel{\hbox{\rlap{\hbox{\lower4pt\hbox{$\sim$}}}\hbox{$<$}}}}
\newcommand{\gsim}{
\mathrel{\hbox{\rlap{\hbox{\lower4pt\hbox{$\sim$}}}\hbox{$>$}}}}

\newcommand{\mueff}{\mu_{\rm eff}}
\renewcommand{\slash}{\displaystyle{\not}}

\newcommand{\ba}{\begin{array}}
\newcommand{\ea}{\end{array}}
\newcommand{\bi}{\begin{itemize}}
\newcommand{\ei}{\end{itemize}}

\newcommand{\N}{\chi^0}

\preprint{
\hbox to \hsize{
\hfill$\vcenter{\hbox{\bf MADPH-06-1473}
                \hbox{\bf hep-ph/0611239}
                \hbox{November 2006}}$}
}

\begin{document}

\title{\vspace*{.75in}
Collider Signatures of Singlet Extended Higgs Sectors}

\author{
Vernon Barger$^1$, Paul Langacker$^{2}$, and Gabe Shaughnessy$^1$}

\affiliation{
$^1$Department of Physics, University of Wisconsin,
Madison, WI 53706 \\
$^2$ School of Natural Sciences, Institute for Advanced Study, 
Einstein Drive, Princeton, NJ 08540
\vspace*{.5in}}

\thispagestyle{empty}

\begin{abstract}
\noindent
We explore the collider signatures of the Higgs sectors in singlet-extended MSSM models.  We find that even with reduced couplings due to singlet mixing, a significant portion of the parameter spaces have a discoverable Higgs via traditional decay modes or via invisible decays (directly to neutralinos or through cascade decays to neutralinos and neutrinos).  For illustrative points in parameter space we give the likelihood of Higgs discovery.  In cases where neither traditional nor invisible modes can discover the Higgs, the neutralino sector may provide evidence for the extended models.
\end{abstract}
\maketitle

\newpage
\section{Introduction}

The Minimal Supersymmetric Standard Model (MSSM) is a leading candidate for beyond the standard model (SM) physics.  The motivation for the MSSM is extensive and includes solutions to the gauge hierarchy problem, the quadratic divergence in the Higgs boson mass, gauge coupling unification, and a viable dark matter candidate.   In the MSSM lagrangian the Higgsino mixing parameter, $\mu$, is the only massive parameter that is SUSY conserving.  Its value sets the scale of the electroweak symmetry breaking in the MSSM and is thus required to be at the electroweak (EW) or TeV scale, though a priori it could be at any value \cite{muproblem}.  

Supersymmetric models with an additional singlet Higgs field address the fine-tuning problem of the MSSM by promoting the $\mu$ parameter to a dynamical field whose vacuum expectation value $\langle S\rangle$ and coupling $\lambda$ determine the effective $\mu$-parameter,
\be
\mu_{\rm eff} = \lambda \langle S\rangle.
\ee
Depending on the symmetry imposed on the theory, a variety of singlet extended models (xMSSM) may be realized.  The models we focus on include the Next-to-Minimal Supersymmetric SM (NMSSM) \cite{NMSSM}, the Nearly-Minimal Supersymmetric SM (nMSSM) \cite{nMSSM, Panagiotakopoulos:2000wp,Menon:2004wv}, and the $U(1)'$-extended MSSM (UMSSM) \cite{UMSSM}, as detailed in Table \ref{tbl:model} with the respective symmetries \footnote{There have also been studies of singlet extensions in a non-supersymmetric context \cite{O'Connell:2006wi}.}.  A Secluded $U(1)'$-extended MSSM (sMSSM) \cite{sMSSM,Han:2004yd} contains three singlets in addition to the standard UMSSM Higgs singlet; this model is equivalent to the nMSSM in the limit that the additional singlet vevs are large, and the trilinear singlet coupling, $\lambda_s$, is small \cite{Barger:2006dh}.  The nMSSM and sMSSM will therefore be referred to together as the n/sMSSM.

The additional CP-even and CP-odd Higgs bosons, associated with the inclusion of the singlet field, yield interesting experimental consequences at colliders.  For recent reviews of these models including their typical Higgs mass spectra and dominant decay modes, see Refs. \cite{Barger:2006dh,Kraml:2006ga}.

\begin{table}[tb]
\caption{Higgs bosons of the MSSM and several of its extensions.  We denote the single CP-odd state in the MSSM and UMSSM by $A_2^0$ for easier comparison with the other models. Possible CP violation, which could induce mixing between CP-even and odd states, is ignored.}
\begin{center}
\label{tbl:model}
\begin{tabular}{|c|lllll|}
\hline
Model:~~& MSSM &NMSSM &{nMSSM}&{UMSSM}&{sMSSM}\\
\hline
Symmetry:~~  & ~~--                             & ~~~$\mathbb Z_3$    &$\mathbb Z^R_5, \mathbb Z^R_7$      & $U(1)'$ & ~~$U(1)'$\\\hline
Extra   &~~--         &       ~~${\kappa\over3} \hat S^3$    &~~$t_F \hat S$& ~~-- &$\lambda_s \hat S_1 \hat S_2 \hat S_3$\\
superpotential term&~~--         &     (cubic)    &(tadpole) &  & (trilinear)\\\hline
& $H_1, A_2$      &     $H_1, A_1$ 	&$H_1, A_1$& $H_1, A_2$ &$H_1, A_1$\\
&   $H_2$	 	&      $H_2, A_2$	&$H_2, A_2$& $H_2$          &$H_2, A_2$\\
CP even, odd&        &      $H_3$ 	  &$H_3$	& $H_3$	 &$H_3, A_3$\\
Higgs bosons&        &        		  &		&  		 &$H_4, A_4$\\
&        &        		  &		&  		 &$H_5$\\
&        &        		  &		&  		 &$H_6$\\
\hline
\end{tabular}
\end{center}
\end{table}

\begin{figure}[htbp]
\begin{center}
\includegraphics[angle=-90,width=0.49\textwidth]{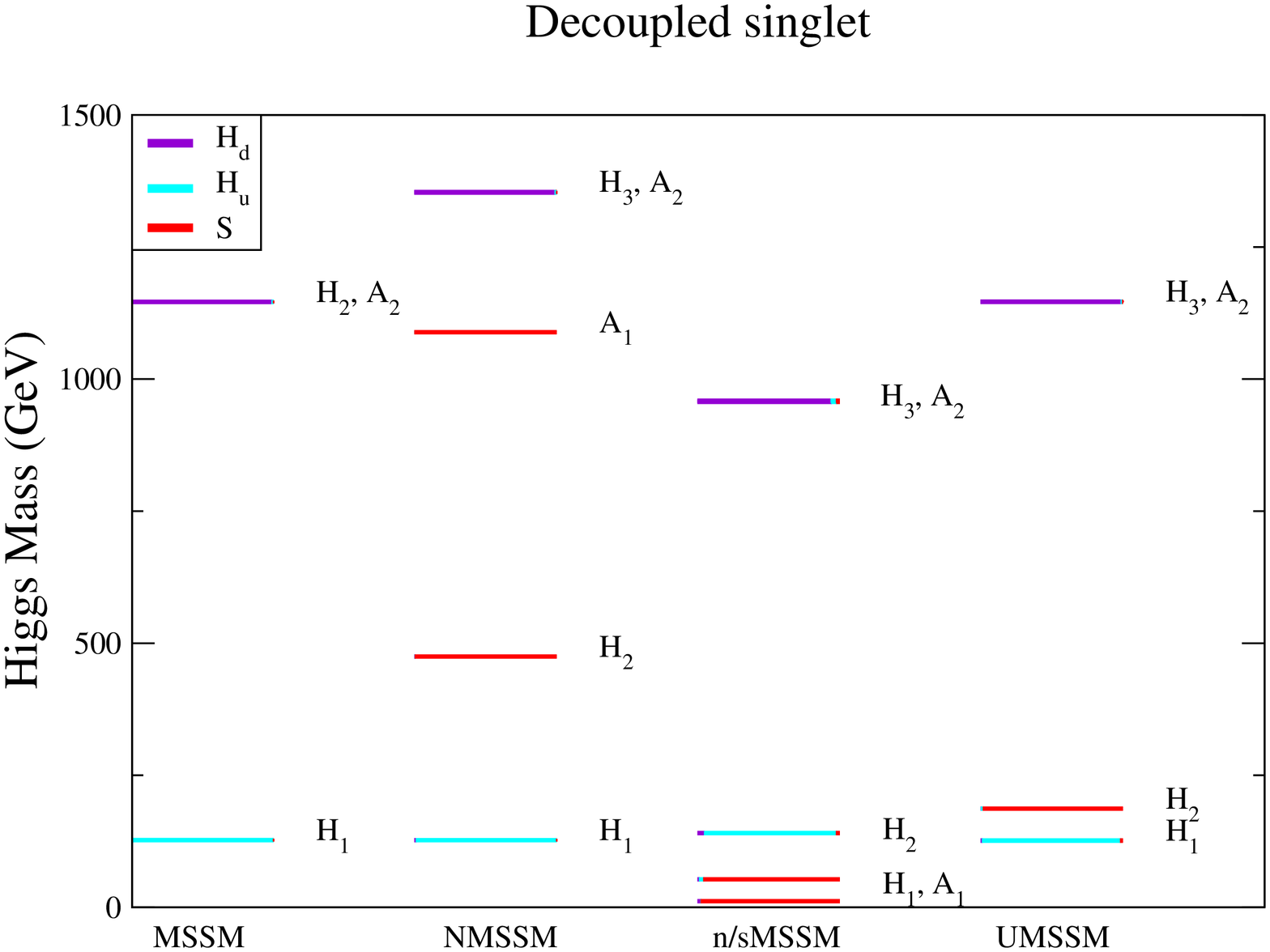}
\includegraphics[angle=-90,width=0.49\textwidth]{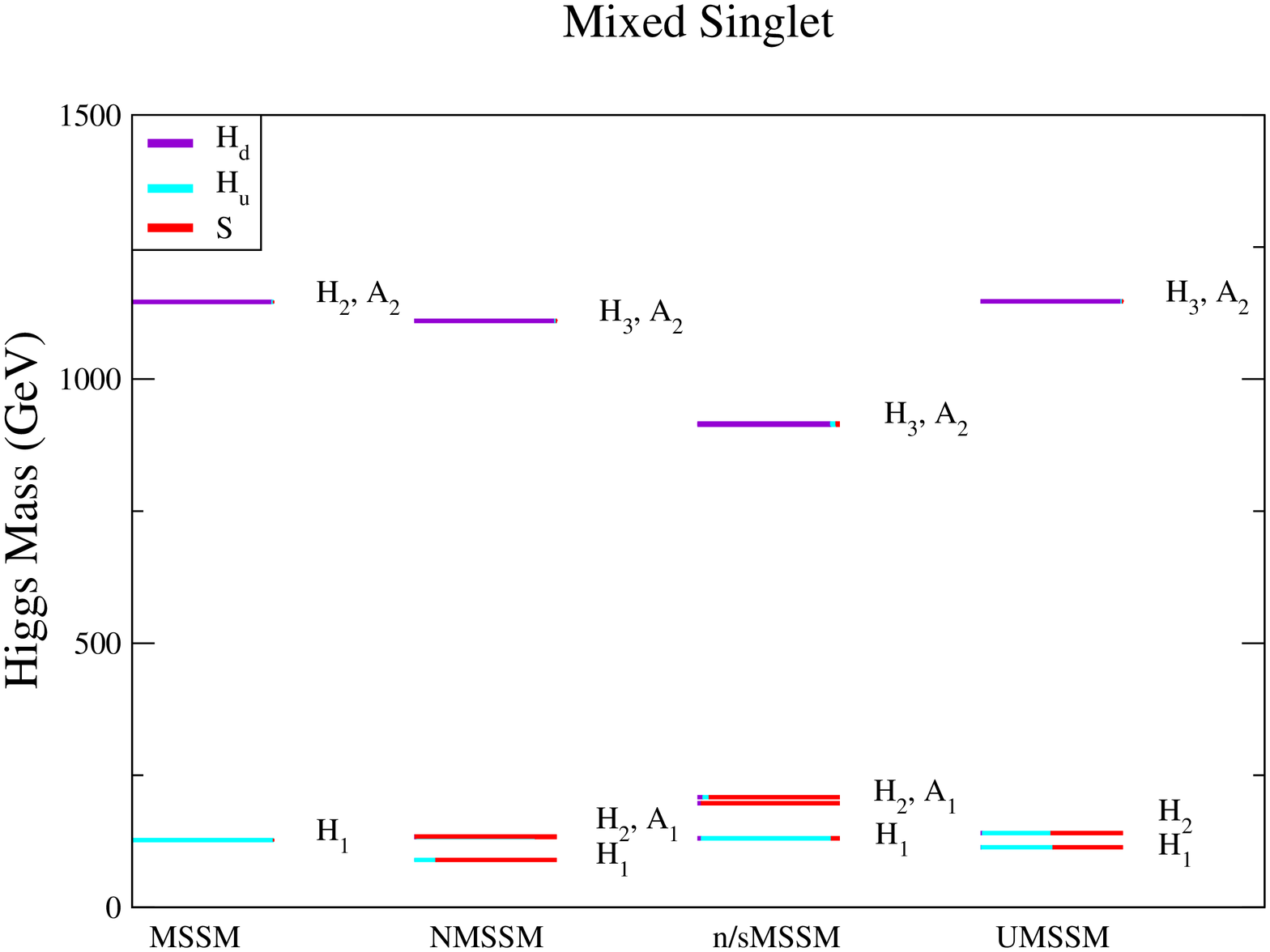}
(a)\hspace{0.48\textwidth}(b)
\caption{Illustrative Higgs composition $(H_d, H_u, S)$ for the models in (a) a decoupled singlet scenario and (b) a strongly mixed singlet scenario.  In the decoupled scenario, the extended model has a spectrum similar to that of the MSSM, but contains an additional singlet Higgs that is heavy in the NMSSM and UMSSM and light in the n/sMSSM.  Parameters used for this illustration are $\tan \beta = 10$, $s=800$ GeV, $\mueff = 130$ GeV, $M_2 = 250$ GeV, $A_{\lambda} = 1$ TeV and $\theta_{E_6} = 0.67$ for the UMSSM.  The n/sMSSM parameter values are $\tan \beta = 5$, $s=400$ GeV, $\mueff = 210$ GeV and $M_2 = -140$ GeV.  For (a) $\kappa=0.7$ and $A_{\kappa}=-1$ TeV in the NMSSM and $t_F = -0.025$ TeV$^2$ and $t_S =-0.00125$ TeV$^3$ in the n/sMSSM; for (b) $\kappa=-0.11$ and $A_{\kappa}=100$ GeV in the NMSSM and $t_F =  -0.0625$ TeV$^2$  and $t_S = -0.0125$ TeV$^3$ in the n/sMSSM and $s=550$ GeV in the UMSSM. $A_{\lambda}, A_{\kappa}$ and $t_S$ are respectively the soft parameters associated with $\lambda, \kappa$ and $t_F$. }
\label{fig:illust}
\end{center}
\end{figure}

To illustrate the Higgs sector of the extended models in the cases in which the lightest Higgs is either decoupled or strongly mixed with the MSSM Higgs boson, we present in Fig. \ref{fig:illust} the neutral Higgs mass spectra for particular points in parameter space.  With sufficient mixing, the lightest Higgs boson can evade the current LEP bound on the SM Higgs mass in these models \cite{Barger:2006dh,Han:2004yd,Dermisek:2005gg}.  Also, singlet interactions may increase the Higgs mass beyond the MSSM theoretical limit \cite{Barger:2006dh,Kraml:2006ga,Barbieri:2006bg}.  Later we discuss the prospects that these Higgs states can be observed at the LHC by direct and indirect searches.

In Section \ref{sect:hadcoll} we consider Higgs signals unique to these models at the Large Hadron Collider (LHC), including the expected signal significances at the ATLAS and CMS detectors.  In Section \ref{sect:invHiggs}, we discuss the observability of an invisibly decaying Higgs that is not uncommon in these models.  In Section \ref{sect:coupmeas}, Higgs coupling measurements are explored as a means to further distinguish the extended singlet models.  Finally, in Section \ref{sect:concl}, we provide concluding remarks.

\section{Signals at Hadron Colliders}\label{sect:hadcoll}

The production of the Higgs bosons at hadron colliders depends on the masses and couplings to SM particles. The couplings of the Higgs bosons to gauge bosons in the singlet models,
\be
\xi_{VVH_i} =  R_{+}^{i1}\cos \beta+R_{+}^{i2}\sin\beta,
\label{eq:coupg}
\ee
 are shared by the three CP-even Higgs bosons.  Here $\xi_{xy H_i}=g_{xyH_i}/g_{xyh_{SM}}$ is the trilinear coupling of fields $x, y$ to the Higgs mass eigenstate $H_i$ in the singlet extended models relative to the corresponding SM coupling.  The matrix, $R_{+}^{ij}$, rotates the Higgs fields from the $\{H^0_d,H^0_u,S\}$ interaction basis to the mass basis (see Eq. (26) of Ref \cite{Barger:2006dh}).  This creates a complementarity among the Higgs coupling to gauge bosons similar to that found in the MSSM \cite{Baer:2004rs}.  The Yukawa couplings
\be
\xi_{ddH_i} = {R_{+}^{i1}\over\cos\beta},\quad \xi_{uuH_i}={R_{+}^{i2}\over\sin\beta},
\label{eq:coupy}
\ee
satisfy the sum rules
\bea
\sum_i \xi_{ff H_i}^2=\left\{ \begin{array}{ccc}
	1+ \cot^2\beta & & \text{up-type quarks}\\
	1+ \tan^2\beta &  &\text{down-type quarks}\\
	\end{array}\right. ,
\eea
which thereby bound the sum of Yukawa coupling squares to a given Higgs state.  Although the Yukawa couplings are bounded by the sum rules, they may individually be significantly larger than the corresponding SM couplings; this can be important in Higgs production and decay.

\begin{figure}[htbp]
\begin{center}\vspace{-.25in}
\includegraphics[angle=-90,width=0.49\textwidth]{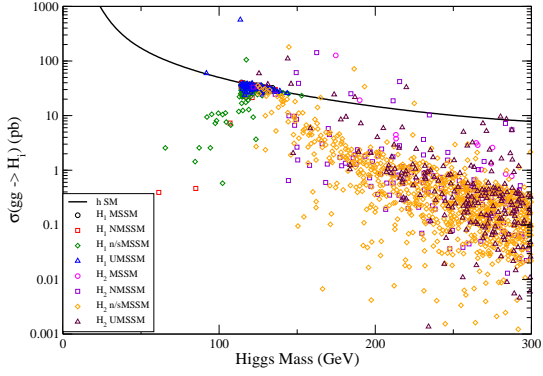}
\includegraphics[angle=-90,width=0.49\textwidth]{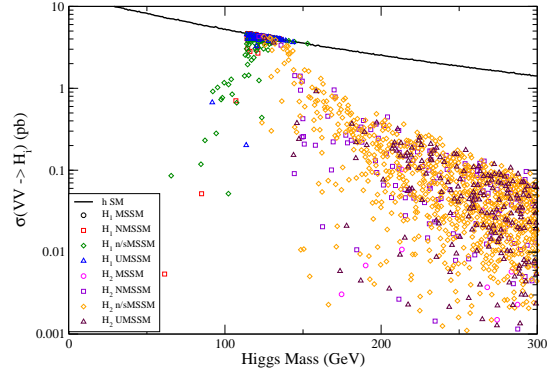}
(a)\hspace{0.48\textwidth}(b)\vspace{-.25in}
\includegraphics[angle=-90,width=0.49\textwidth]{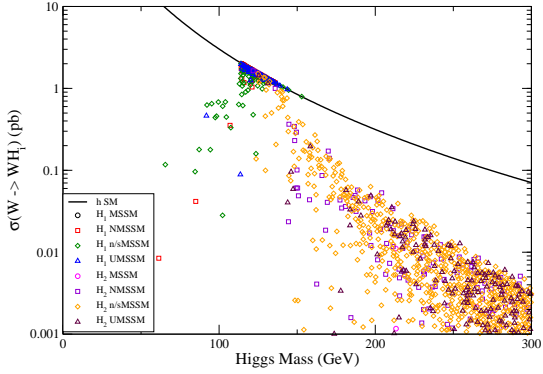}
\includegraphics[angle=-90,width=0.49\textwidth]{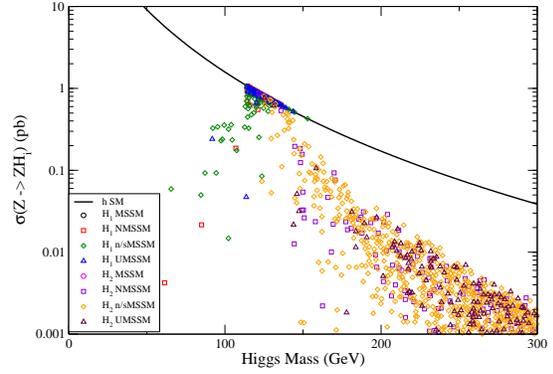}
(c)\hspace{0.48\textwidth}(d)\vspace{-.25in}
\includegraphics[angle=-90,width=0.49\textwidth]{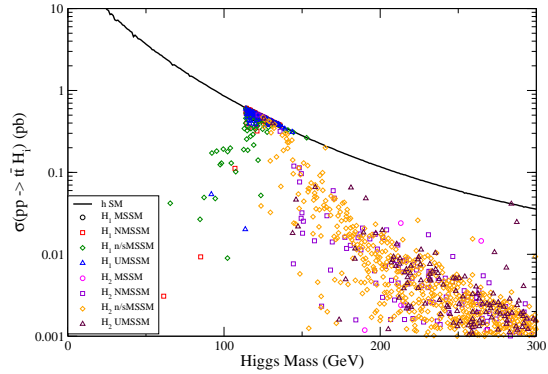}
\hspace{0.48\textwidth}(e)
\caption{Production cross section of the lightest Higgs in the MSSM and extended MSSM models at the LHC via (a) gluon fusion, (b) Weak Boson Fusion (WBF), (c) Higgstrahlung of a $W$ boson, (d) Higgstrahlung from a $Z$ boson and (e) associated production with top pairs. Most of the MSSM $H_1$ points lie close to the SM curves because they are in the MSSM decoupling limit.}
\label{fig:hprod}
\end{center}
\end{figure}

We generate the Higgs spectra (and associated neutralino spectra) of the models and apply experimental constraints following Refs. \cite{Barger:2006dh,Barger:2006kt} \footnote{The scan of the n/sMSSM is slightly different from Ref. \cite{Barger:2006dh} in that a uniformly random scan over the parameters $t_S$ and $t_F$ is performed here whereas the scan of Ref. \cite{Barger:2006dh} concentrates the scan near $|t_S,t_F| \sim 0$ to accentuate the light Higgs scenarios.}.  By appropriate changes in the Higgs boson couplings to the relevant particles as in Eqs. (\ref{eq:coupg},\ref{eq:coupy}), we calculate the production cross sections via the SM codes of Ref. \cite{ref:hprod}; these codes include the fusion subprocesses $gg \to H_i$, $VV \to H_i$ at NLO, the associated production modes $ V H_i$ at NLO, and $ t \bar t H_i$ at LO.  The production cross sections are shown \footnote{The density of overdense regions of points generated has been reduced in the plots for clarity.} in Fig. \ref{fig:hprod}.   Precision electroweak measurements prefer a heavy SUSY sector \cite{Heinemeyer:2006px}, so we assume that the scalar quarks in the gluon fusion loops are heavy (all soft masses are at 1 TeV) and thus approximately decoupled; the effects of stop mixing changes the Higgs production rate by up to 10\% in our scan with respect to the SM depending on the mixing parameter $X_t = A_t - \mueff \cot \beta$.  

For the lightest Higgs boson in the MSSM a majority of the generated cross section points are close to the SM curve.  The cross sections at low Higgs masses ($\lesssim 100$ GeV) are kinematically enhanced but suppressed by the coupling and are typically 2 orders of magnitude smaller than the corresponding SM cross section.  The production rates with Higgs masses in the range $150\text{ GeV} \lesssim M_{H_i} \lesssim 300\text{ GeV}$ are also typically a few orders of magnitude lower than the SM result.  This is typically due to the large singlet component of that Higgs eigenstate, but can also arise from the MSSM-like Higgs that weakly couples to SM fields.

Over most of the range of Higgs masses, gluon fusion is the dominant production subprocess.  The cross section can be larger than in the SM, as shown in Fig. \ref{fig:hprod}a, due to larger Yukawa couplings.  Production modes via vector boson fusion are subdominant but yield cleaner experimental signals.  The dominant decays of Higgs bosons of intermediate mass $120-150$ GeV, are $H_i \to b \bar b$ and the silver channel $H_i \to W W \to l \nu l \nu$.  However, these branching fractions may be small compared to the SM due to a large decay rate to neutralinos or to light CP-odd pairs \footnote{Decays to light CP-odd pairs that subsequently decay to $b \bar b$ can lead to additional $b$ pairs.  However, if the CP-odd Higgs is very light ($M_{A}\lesssim 10$ GeV), the Higgs may not be directly observable \cite{Dermisek:2005gg}.}  \cite{Barger:2006dh}.  Additional signals unique to singlet models include associated chargino pair production in association with a light CP-odd Higgs boson, decaying to a lepton and photon pair with large missing energy \cite{Arhrib:2006sx}.

\section{Significance of Higgs signals}\label{sect:signif}

A SM Higgs boson can be discovered at the LHC through a variety of channels \cite{ref:cmstdr,ref:atltdr}.  However, in singlet extended supersymmetric models, the expectation may not be so optimistic inasmuch as singlet mixing or decays to additional light neutralinos may spoil direct observation.

\begin{figure}[htbp]
\begin{center}
\includegraphics[angle=-90,width=0.49\textwidth]{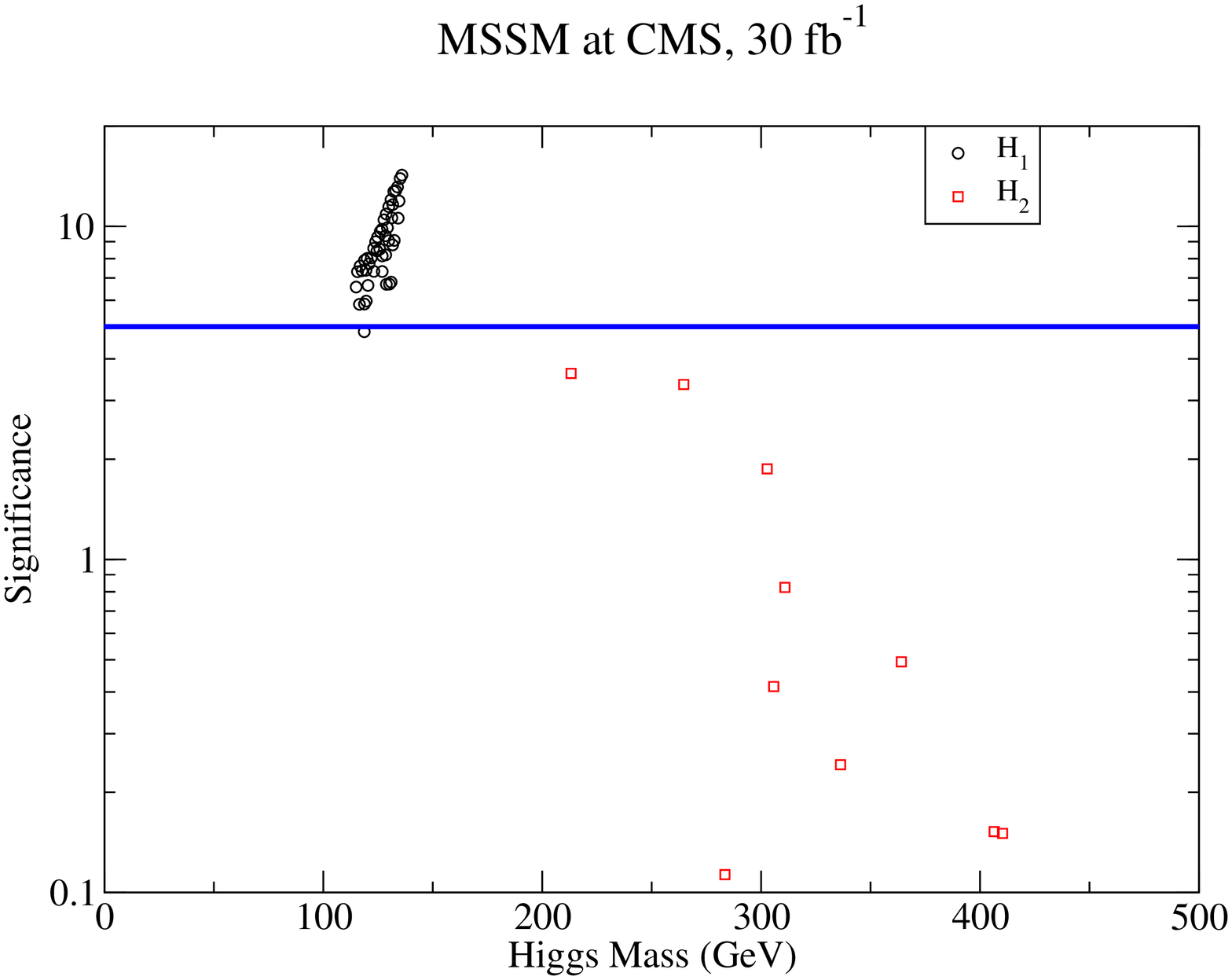}
\includegraphics[angle=-90,width=0.49\textwidth]{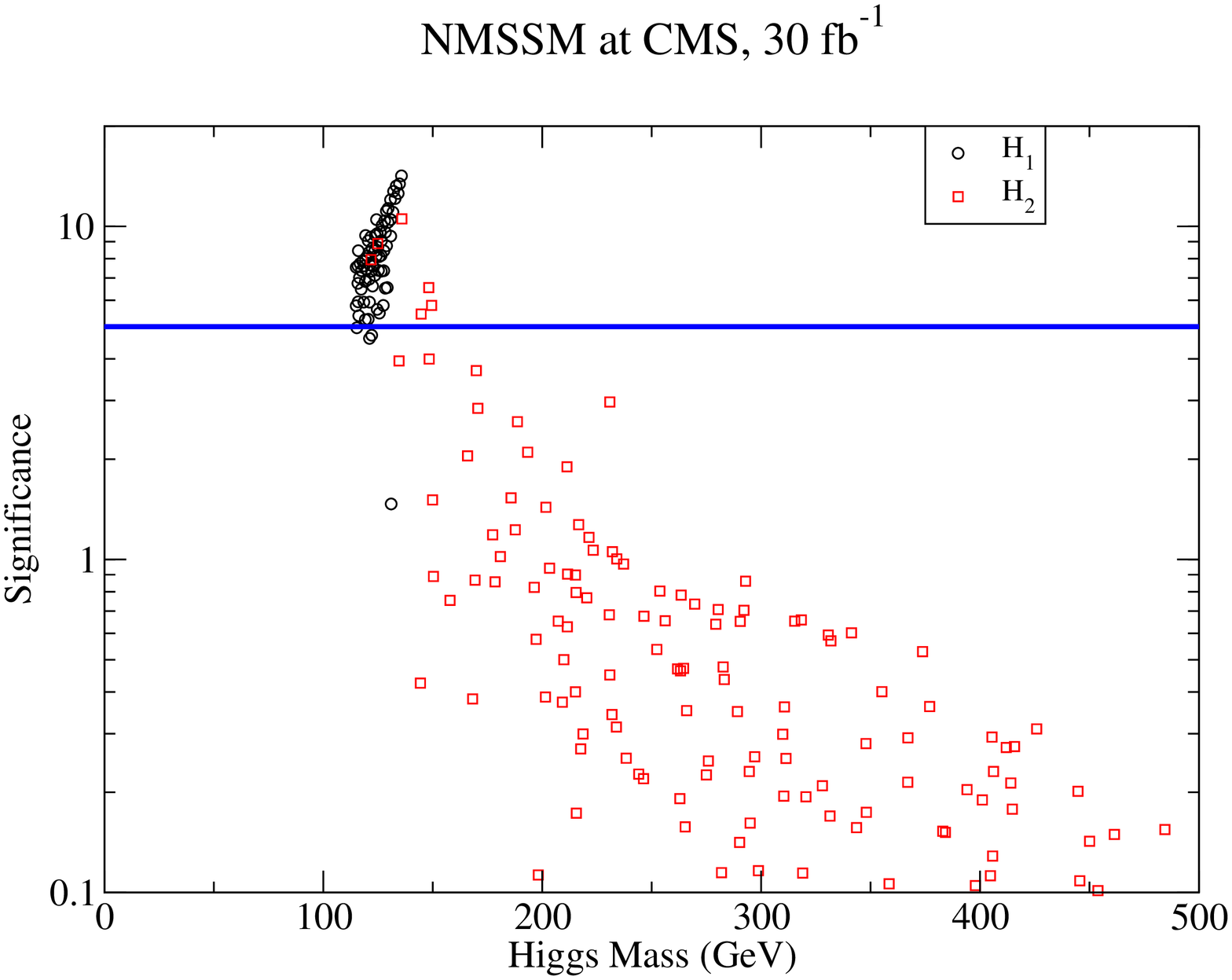}
(a)\hspace{0.48\textwidth}(b)\vspace{-.25in}
\includegraphics[angle=-90,width=0.49\textwidth]{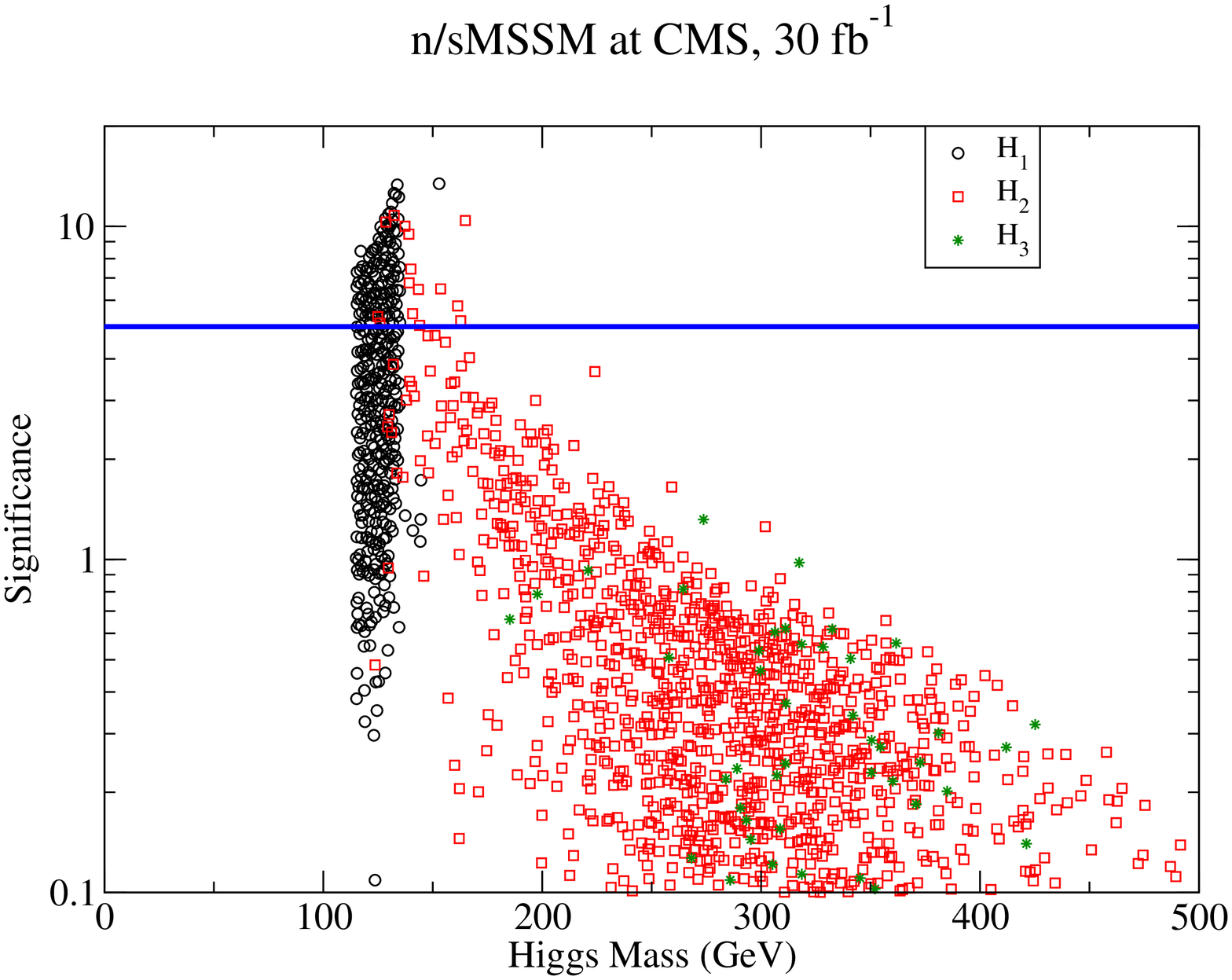}
\includegraphics[angle=-90,width=0.49\textwidth]{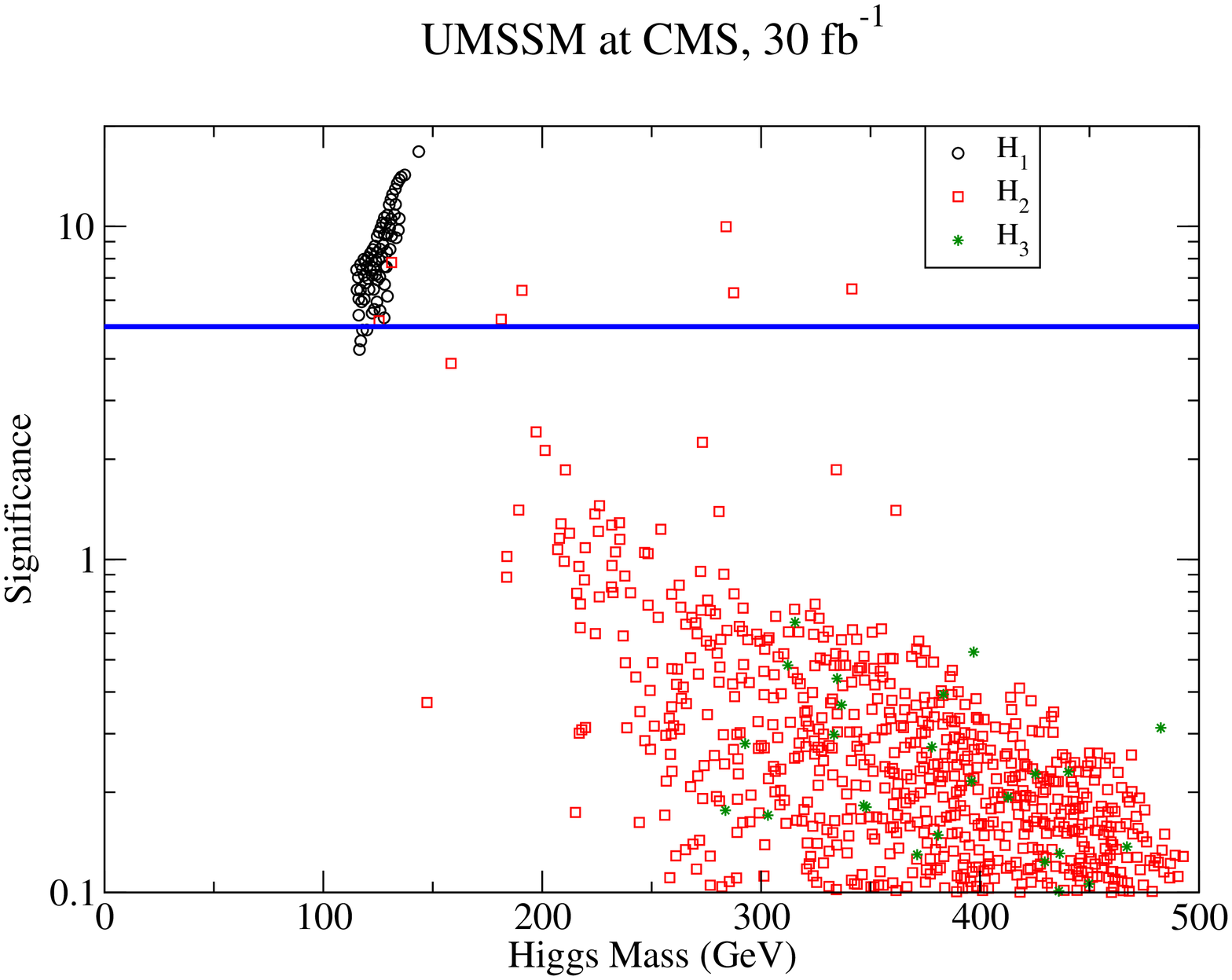}
(c)\hspace{0.48\textwidth}(d)
\caption{Signal significance $S/\sqrt{B}$ at CMS \cite{ref:cmstdr} with 30 fb$^{-1}$ of data in the (a) MSSM (b) NMSSM (c) n/sMSSM (d) UMSSM.  The significance of the lightest Higgs is dominated by $gg\to H$ with $H\to \gamma \gamma$ and $H\to ZZ \to 4l$.  The $H\to ZZ \to 4l$ mode provides the best signal for heavier Higgs masses, even though these Higgs states are typically singlet or MSSM-like and weakly couple to the $Z$ bosons.  No K-factors have been applied as in the ATLAS and CMS analyses.}
\label{fig:cmssig}
\end{center}
\end{figure}

\begin{figure}[htbp]
\begin{center}
\includegraphics[angle=-90,width=0.49\textwidth]{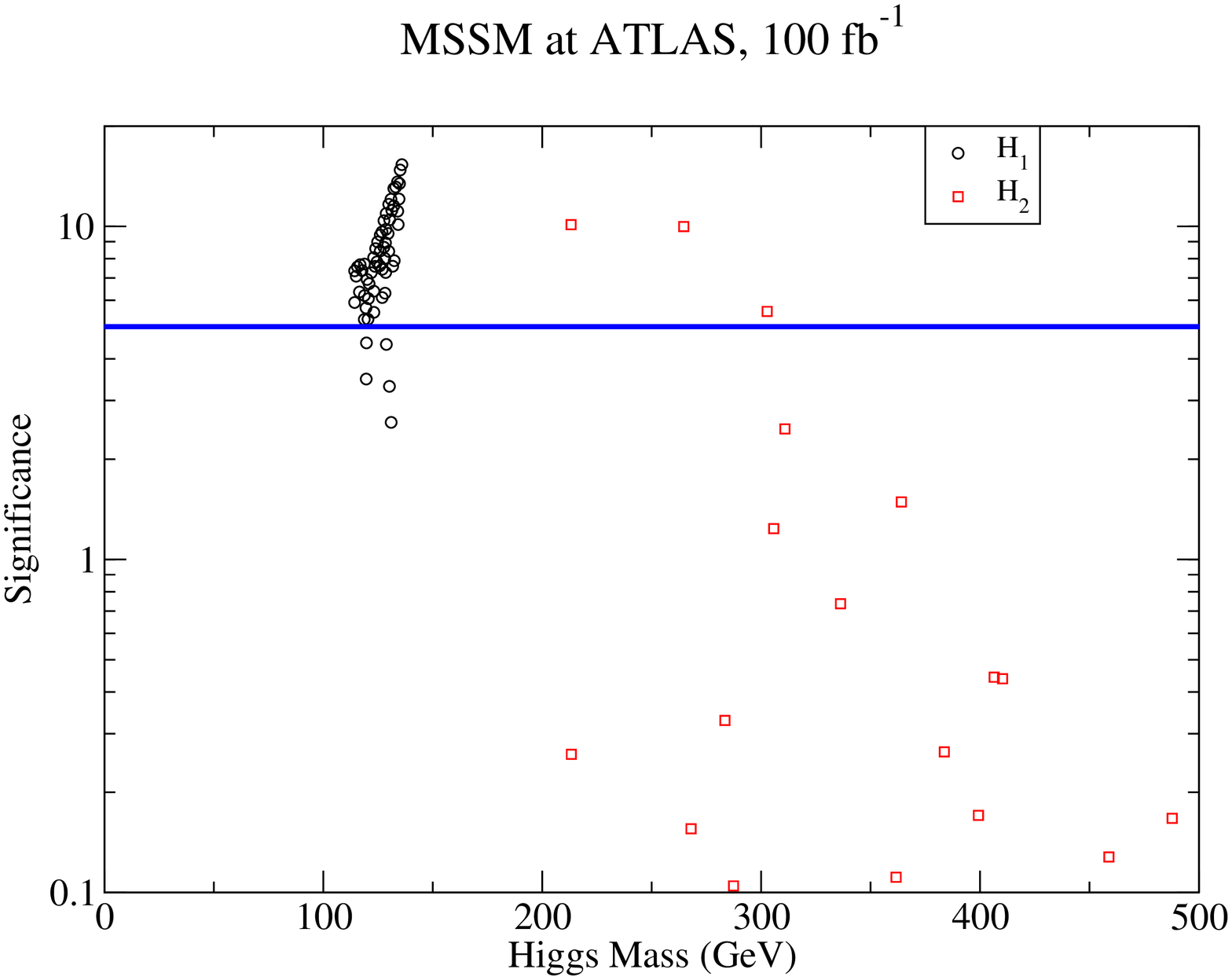}
\includegraphics[angle=-90,width=0.49\textwidth]{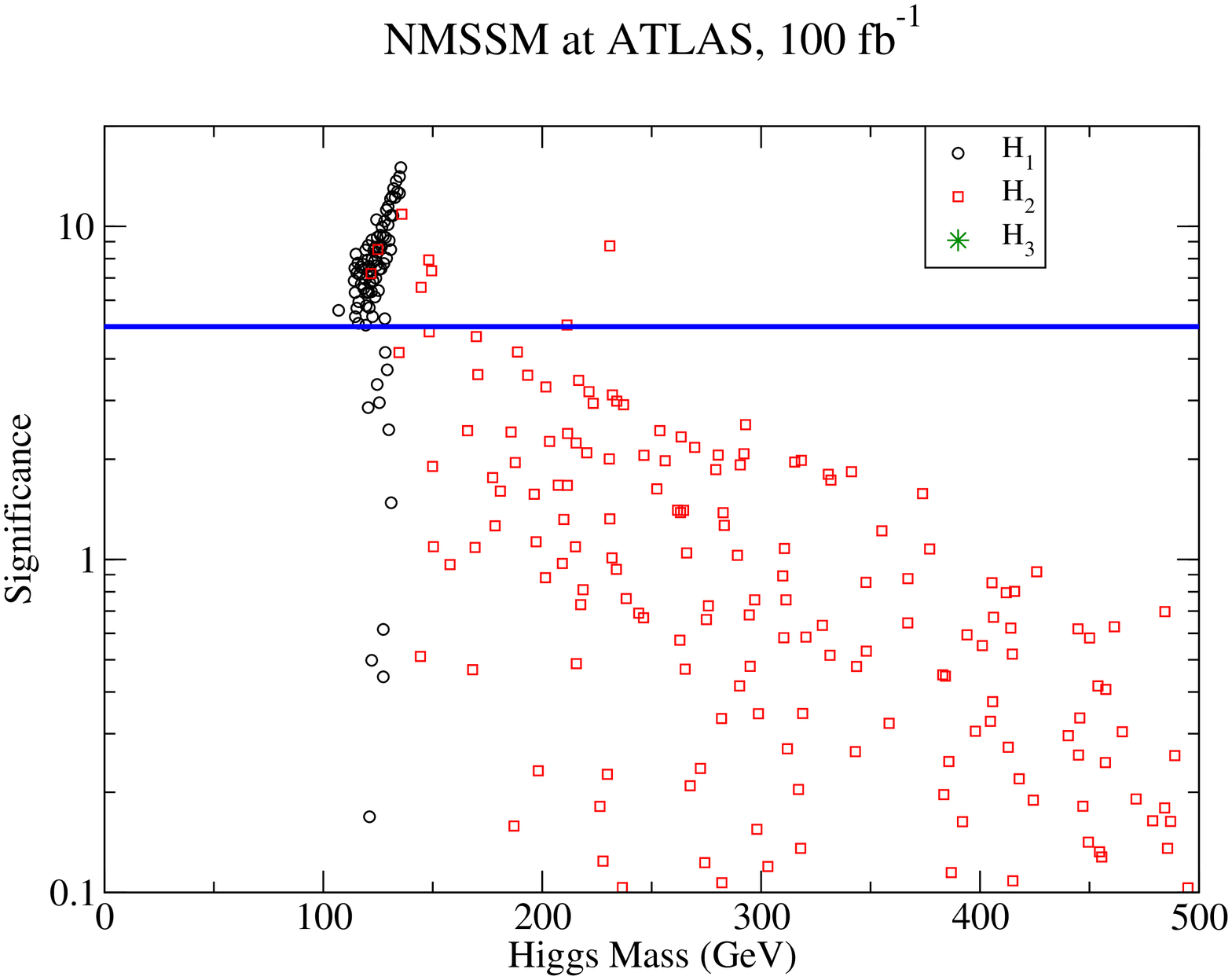}
(a)\hspace{0.48\textwidth}(b)\vspace{-.25in}
\includegraphics[angle=-90,width=0.49\textwidth]{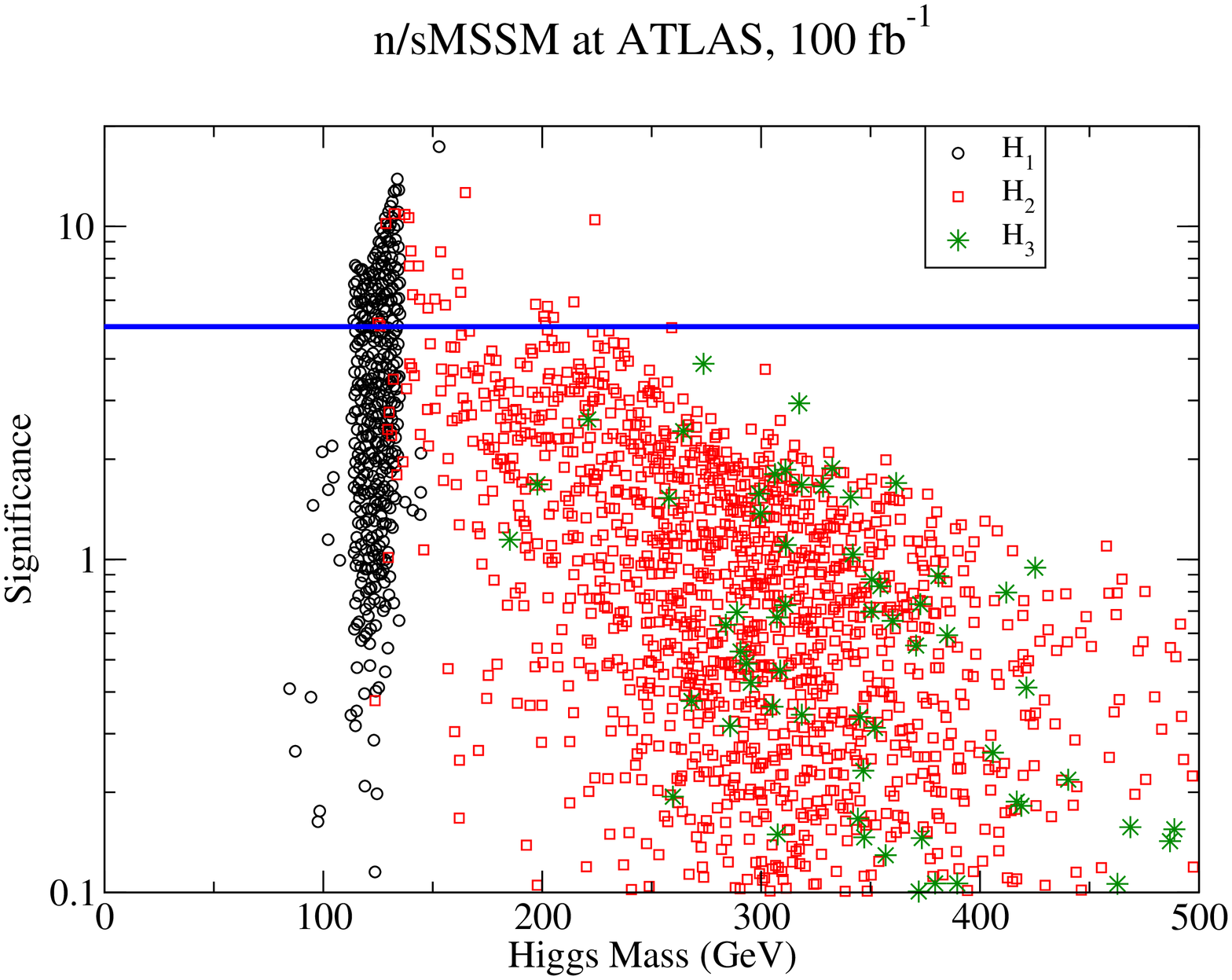}
\includegraphics[angle=-90,width=0.49\textwidth]{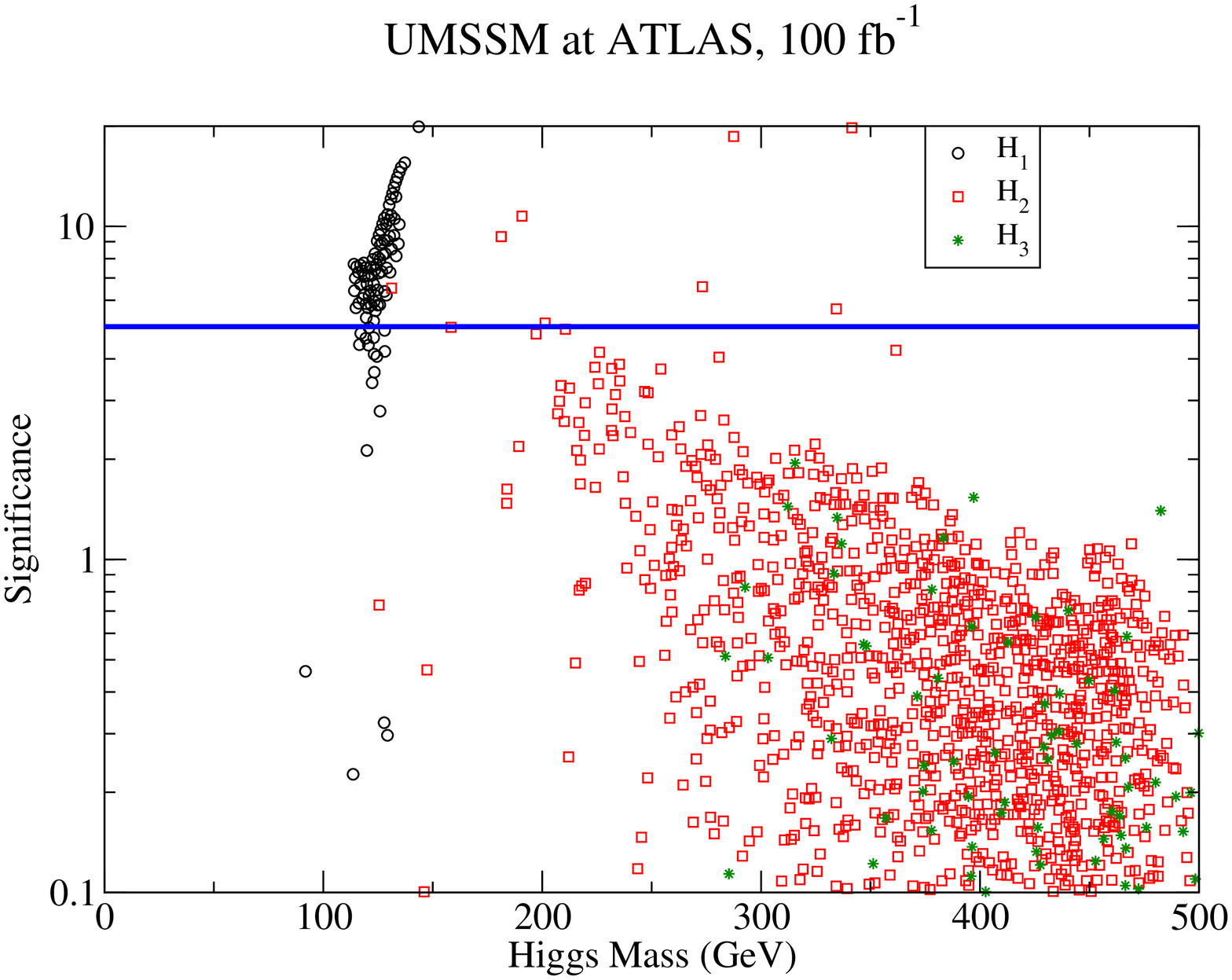}
(c)\hspace{0.48\textwidth}(d)
\caption{Same as in Fig. \ref{fig:cmssig}, except for ATLAS \cite{ref:atltdr} with 100 fb$^{-1}$ of data.}
\label{fig:atlsig}
\end{center}
\end{figure}

The ATLAS and CMS collaborations have evaluated the expected significances for 5 sigma discovery of the SM Higgs boson, with 100 fb$^{-1}$ at ATLAS and 30 fb$^{-1}$ at CMS \cite{ref:cmstdr,ref:atltdr}.  We convert their SM results to determine the prospects of discovery of the Higgs bosons of the singlet extended models.  The total significance for each model, shown in Figs. \ref{fig:cmssig} and \ref{fig:atlsig}, is obtained by summing the significances of the contributing modes in quadrature.  The significances of the individual modes are found by scaling the SM significances with the production couplings relative to the SM and the branching fractions relative to the SM (i.e. for $VV\to H_i\to ZZ$, the scaling is $\xi^2_{VVH_i} {\text{Bf}(H_i \to ZZ)\over \text{Bf}(h_{SM} \to ZZ)}$).  The SM significance evaluations are not available for $M_H \lesssim 110$ GeV where the larger backgrounds make SM Higgs detection at the LHC problematic \footnote{Many points with Higgs masses below the SM Higgs 114 GeV LEP limit are often suppressed below a significance of 0.1 due to the dominant singlet fraction.  In the cases where the lightest Higgs is decoupled and light (typically in the n/sMSSM), the second Higgs state, $H_2$, has a mass in the range of the lightest Higgs in the MSSM and a large significance.}.

The modes contributing to the total significance for both ATLAS and CMS include $gg\to H$ with $H\to\gamma \gamma$, $H\to ZZ\to 4l$, $H\to WW\to l\nu l\nu$.  Additional modes in the analysis of ATLAS are $H\to ZZ \to l l \nu \nu$, $t \bar t H$ with $H\to b \bar b$ and the Higgstrahlung process $H W \to WWW\to l\nu l\nu l\nu$; the modes specific to the CMS analysis are the Weak Boson Fusion (WBF) processes $WW \to H$ with $H\to WW \to l \nu jj$, $H\to \tau \tau \to l+j$ and $H\to \gamma \gamma$.

The most promising discovery channel over most of the Higgs mass range is the golden channel $H_i \to ZZ^* \to 4l$ since it has very low backgrounds.  This channel is expected to permit SM Higgs discovery for Higgs masses $120- 600$ GeV.  The golden channel can also be used to determine the spin and CP properties of the decaying Higgs boson by using the correlations of the angle between the planes of lepton pairs \cite{ref:cpspin}.  

For light Higgs bosons ($m_H \lesssim 120$ GeV) the decay $H\to \gamma \gamma$ has the best significance.  Combining this mode with $H\to ZZ\to 4l$ often yields a total significance above $5\sigma$ for the lightest Higgs boson in the MSSM, NMSSM, and UMSSM.  In these models slightly over half of the generated points in parameter space are above 5$\sigma$ for the $H_1$.  The second lightest Higgs in these models is less likely to be discovered at the LHC, but evidence may be found at $3 \sigma$ for the considered integrated luminosities.  The significance of the $H_2$ in extended models (typically a singlet or MSSM-like and weakly coupled to $Z$ bosons) is typically below the 5$\sigma$ level.  This is evident in  Figs. \ref{fig:cmssig} and \ref{fig:atlsig}.

In some cases the signal of the lightest Higgs does not reach the $5\sigma$ discovery limit due to a dominant invisible decay to stable neutralinos that are undetected except as missing transverse energy, $\slash E_T$ \footnote{A decrease in the total significance may also occur if the Higgs boson decays to heavy neutrinos \cite{Belotsky:2002ym}.}.  When the $H\to \N_1 \N_1$ decay channel  is open, the Higgs boson is generally invisible.  In the MSSM, NMSSM, and UMSSM the invisible decay is usually kinematically inaccessible \cite{Barger:2006dh} in our parameter scan, which has a lower limit on $m_{\N_1}$ of 53 GeV (half the chargino mass bound \footnote{We assume gaugino mass unification.}), but the invisible decay is potentially relevant for $M_H > 106$ GeV.  Invisible decays are often dominant in the n/sMSSM where the lightest neutralino mass is typically lighter than 50 GeV \cite{Menon:2004wv,Barger:2006dh,Barger:2005hb,Barger:2006kt}.  Therefore, using traditional searches the discovery of the $H_1$ is unlikely in the n/sMSSM.  However, indirect discovery of an invisibly decaying Higgs is still possible, as discussed in Section \ref{sect:invHiggs}.

Since the extended models with a sufficiently decoupled singlet may mimic the MSSM, discerning which underlying model describes the Higgs sector is a difficult experimental challenge.  Fortunately, there are complementary alternative avenues such as an excess in multilepton events due to additional steps in the cascade decays of neutralinos and charginos \cite{Barger:2006kt}, or the discovery of a $Z'$ in the UMSSM \footnote{There are cases where the indirect detection rates of neutralino dark matter can be enhanced in these models by a light CP-odd Higgs boson \cite{Ferrer:2006hy}.}.  The discovery of a Higgs with mass greater than the theoretical upper limit on the lightest Higgs in the MSSM could indicate a singlet Higgs model \cite{Barger:2006dh}.  Alternatively, discovery of a Higgs with mass less than the LEP bound on the MSSM Higgs mass of 93 GeV would also be evidence that the Higgs sector is not that of the MSSM \footnote{Additionally, an analysis of the Higgs signal at LEP in a model independent way places a lower bound on the Higgs mass at 81 GeV \cite{Abbiendi:2002qp}.  However, this analysis assumes SM strength production and is relaxed significantly if the Higgs is dominantly singlet.}.

\section{Observing an Invisibly decaying Higgs}
\label{sect:invHiggs}

A Higgs boson that decays invisibly can be indirectly inferred by making appropriate cuts on the kinematic distributions of the forward jets in WBF \cite{Eboli:2000ze}.  QCD and electroweak backgrounds due to  $Z jj$ with $Z\to \nu \nu$ and $W^\pm jj$ (with $W\to l \nu$ with the lepton unidentified) give the dominant 2 jet plus missing energy signal. QCD $3j$ production also contributes to the background.  The QCD backgrounds can be significantly reduced with a  $\slash p_T> 100$ GeV cut.  Cuts on the forward jets, $\phi_{jj}<1.0$ and $|\Delta \eta_{jj}|>4.4$, substantially reduce the contribution of back-to-back jets in the $Wjj$ and $Zjj$ backgrounds.  The ratio $R_1 = {\int^1_0 {d\sigma\over d \phi_{jj} }d\phi_{jj}\over \int^\pi_0 {d\sigma\over d \phi_{jj} }d\phi_{jj} }$ can be used to infer the invisible Higgs decay \cite{Eboli:2000ze}.  

\begin{figure}[htbp]
\begin{center}
\includegraphics[angle=-90,width=0.49\textwidth]{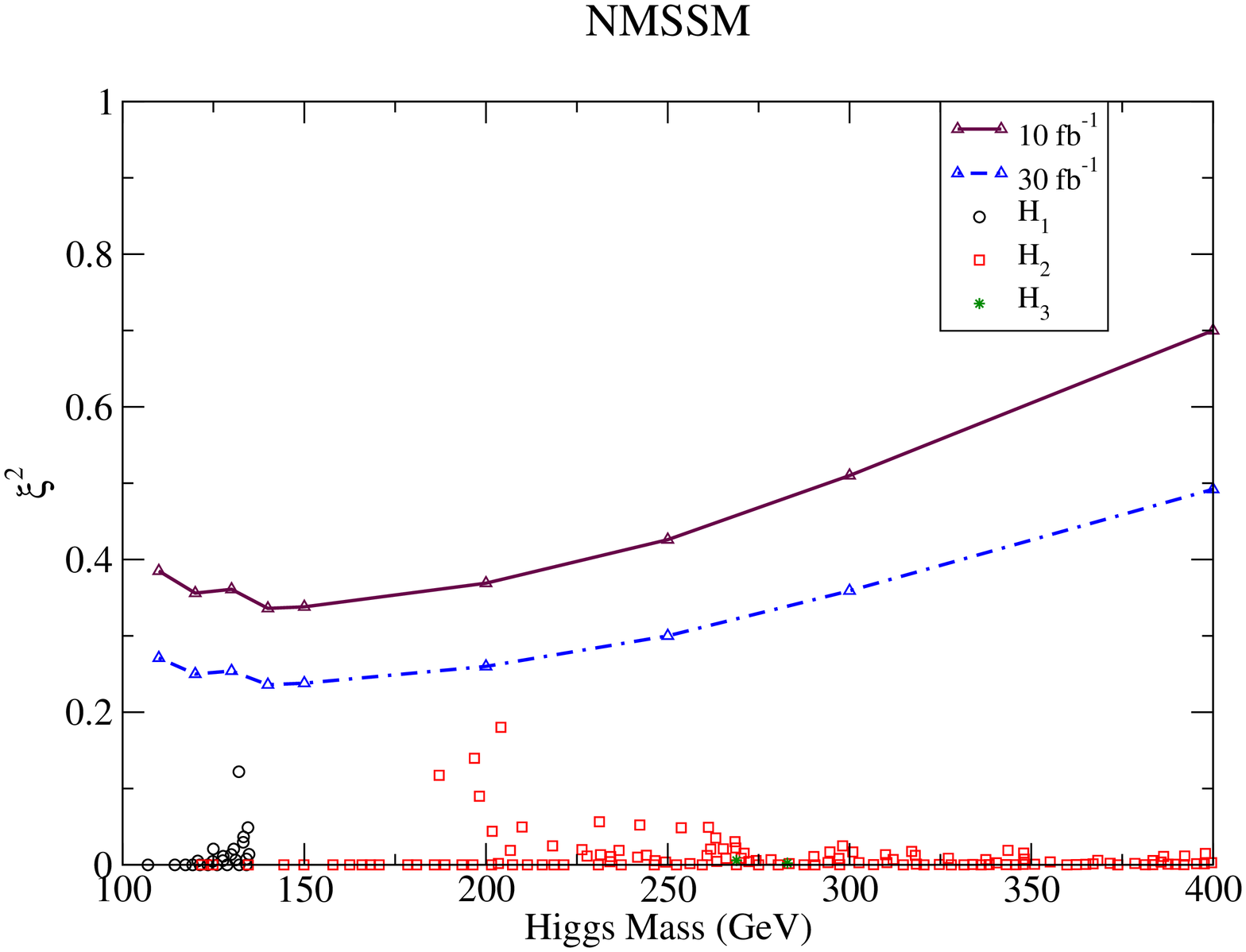}
\includegraphics[angle=-90,width=0.49\textwidth]{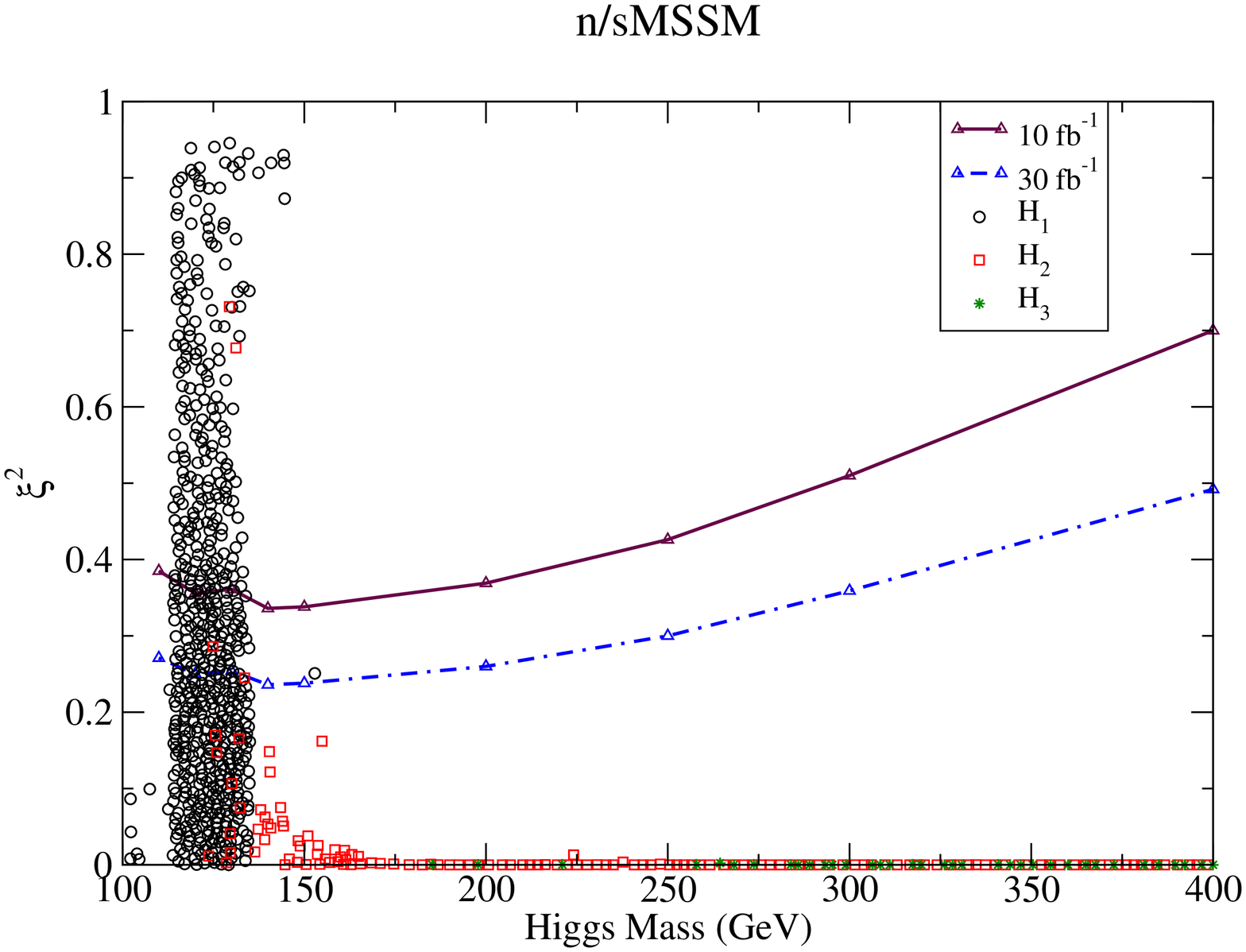}
(a)\hspace{0.48\textwidth}(b)\vspace{-.25in}
\includegraphics[angle=-90,width=0.49\textwidth]{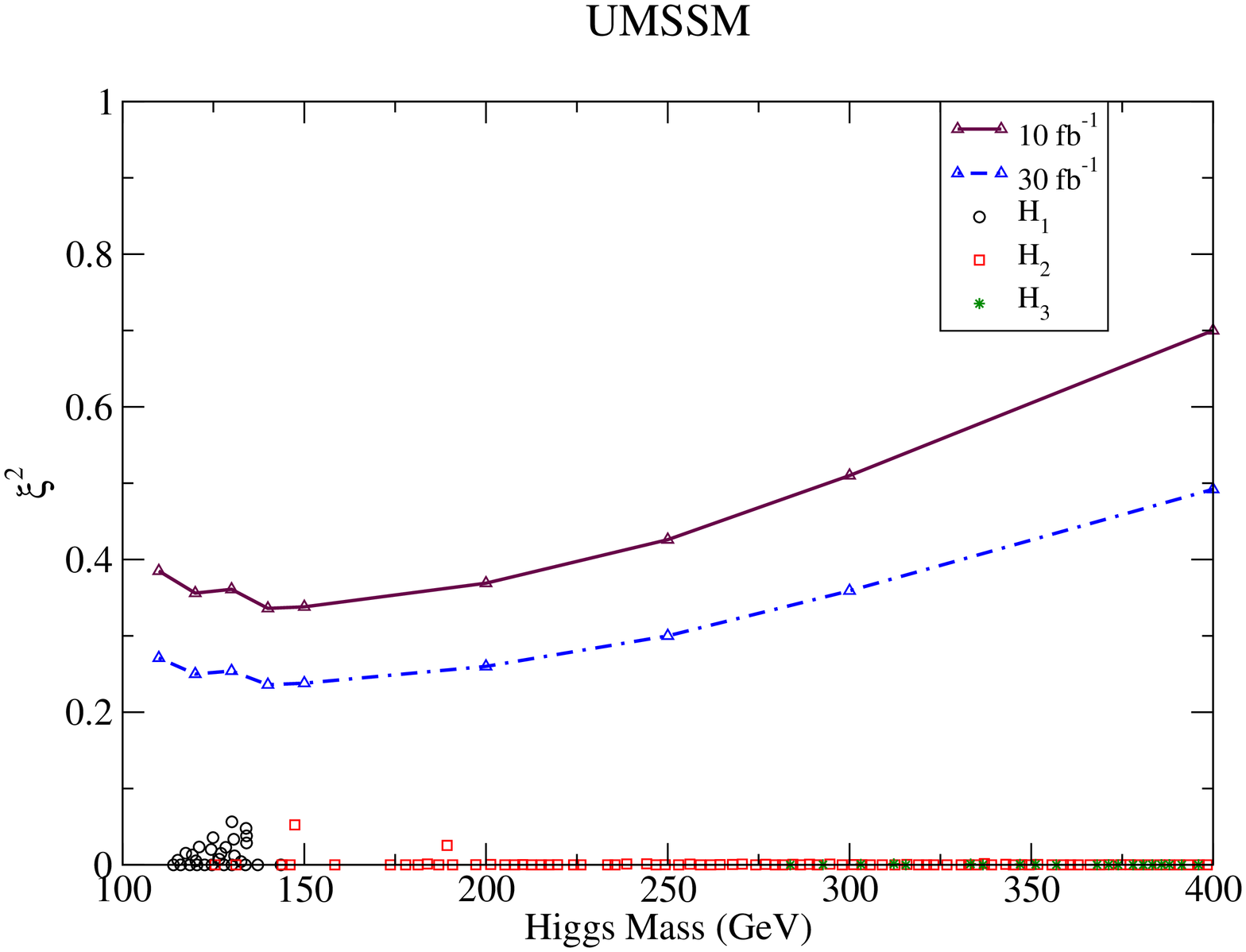}
\hspace{0.48\textwidth}(c)
\caption{Sensitivity of the ATLAS detector to the quantity $\xi_i^2$ of Eq. \ref{eq:invfrac} for invisibly decaying Higgs bosons.  The solid maroon curve indicates the expected sensitivity with 10 fb$^{-1}$ of data while the dashed blue curve represents 30 fb$^{-1}$ of data.  The points are predictions of the NMSSM, n/sMSSM and UMSSM models as labeled.}
\label{fig:invHiggs}
\end{center}
\end{figure}

The quantity
\be
\xi_i^2\equiv BF(H_i\to inv.)\xi_{VVH_i}^2,
\label{eq:invfrac}
\ee
parameterizes the relative amount of invisible Higgs decays in WBF for a given Higgs mass.  Expected sensitivity limits for the observation of an invisible Higgs via WBF are shown for ATLAS with 10 (30) fb$^{-1}$ of data in Fig. \ref{fig:invHiggs} as a solid (dashed) curve \cite{ref:atlinvHiggs}.  The points in the figure show the values of $\xi^2$ for the NMSSM, n/sMSSM and UMSSM, where it is assumed the invisible Higgs signal is due to decays to neutralinos.  

In addition to direct decay to the lightest neutralino, we consider the invisible decays via other neutralino states such as $H_i\to \N_2 \N_1\to 2 \N_1 \nu \bar \nu$ and other cascades involving $\N_2$ or $\N_3$ when accessible while including two and three body decays of the neutralino.  The rate of $H_i \to 2 \N_1 \nu \bar \nu$ is dominated by the process involving the $Z$ boson since the slepton masses are assumed to be 1 TeV.  Also, we include the heavier Higgs boson cascading to invisible modes, $H_2 \to H_1 H_1 \to 4 \N_1$, which can be the dominant invisible channel in the n/sMSSM.  The dilepton decays of the $Z$ boson in the cascade $H_i\to \N_2 \N_1 \to 2 \N_1 Z$ may also contribute to the invisible signal when the leptons are below the $p_T$ acceptance cuts (or may be directly observable via the leptons).  However, relic density constraints place a loose lower bound on the $\N_1$ mass of $m_{\N_1} \gtrsim 30$ GeV \cite{Barger:2005hb}, limiting the kinematic accessibility of this channel.  

In many cases, the Higgs decay to $\N_2 \N_1$ is in fact stronger than to a $\N_1$ pair.  Since the lightest neutralino is dominantly singlino, it couples to the Higgsino part of the second neutralino for an MSSM-like Higgs; the singlino pair couples weakly to a MSSM-like Higgs (see Ref. \cite{Barger:2006kt} for the coupling) \footnote{The NMSSM allows a larger coupling between two singlinos and the lightest Higgs provided the Higgs is dominantly singlet.  However, the decay is not typically kinematically allowed in the NMSSM.}.

The n/sMSSM Higgs is most likely to be discovered from invisible cascade decays, due to the light $\N_1$ mass.  Although the NMSSM and UMSSM both have a fraction of Higgs decays that are invisible, large values of $\xi_i^2$ do not dominate the parameter space of these models, and all of the points fall below the 30 fb$^{-1}$ sensitivity reach of ATLAS.  In the MSSM the invisible branching fraction is very small.  

Another method for invisible Higgs discovery compares the signals from WBF and Higgstrahlung \cite{Davoudiasl:2004aj}.  The WBF signal is found by the above method.  The signal from $Z H_i$ production, with $Z\to l^+ l^-$, can be isolated quite well from the $WW$ background by requiring the invariant mass of the lepton pair to be close to the $Z$ mass.  The background from $ZZ$ production with one $Z\to \nu \bar\nu$ can be reduced with a cut $\slash p_T \gtrsim 75$ GeV.  Since the Higgstrahlung cross section is more sensitive to the Higgs mass than WBF due to the $s$-channel suppression, the ratio of the rates of these processes yields a strong kinematic dependence on $M_H$.  From the ratio, an uncertainty of ${\cal O}(20\text{ GeV})$ can be placed on a Higgs boson of intermediate mass (i.e. $M_h = 120 - 160$) with 100 fb$^{-1}$ of data at the LHC, assuming production with SM strength and fully invisible decays \cite{Davoudiasl:2004aj}.  A reduced $VVH_i$ coupling will reduce the accuracy in determining the Higgs mass by this means.  However, most of the Higgs bosons of the MSSM and extended models that are in the above mass range have an ${\cal O}(1)$ coupling to gauge boson pairs, making the strength of the signal sensitive to the invisible branching fraction.

\begin{figure}[htbp]
\begin{center}
\includegraphics[angle=-90,width=0.49\textwidth]{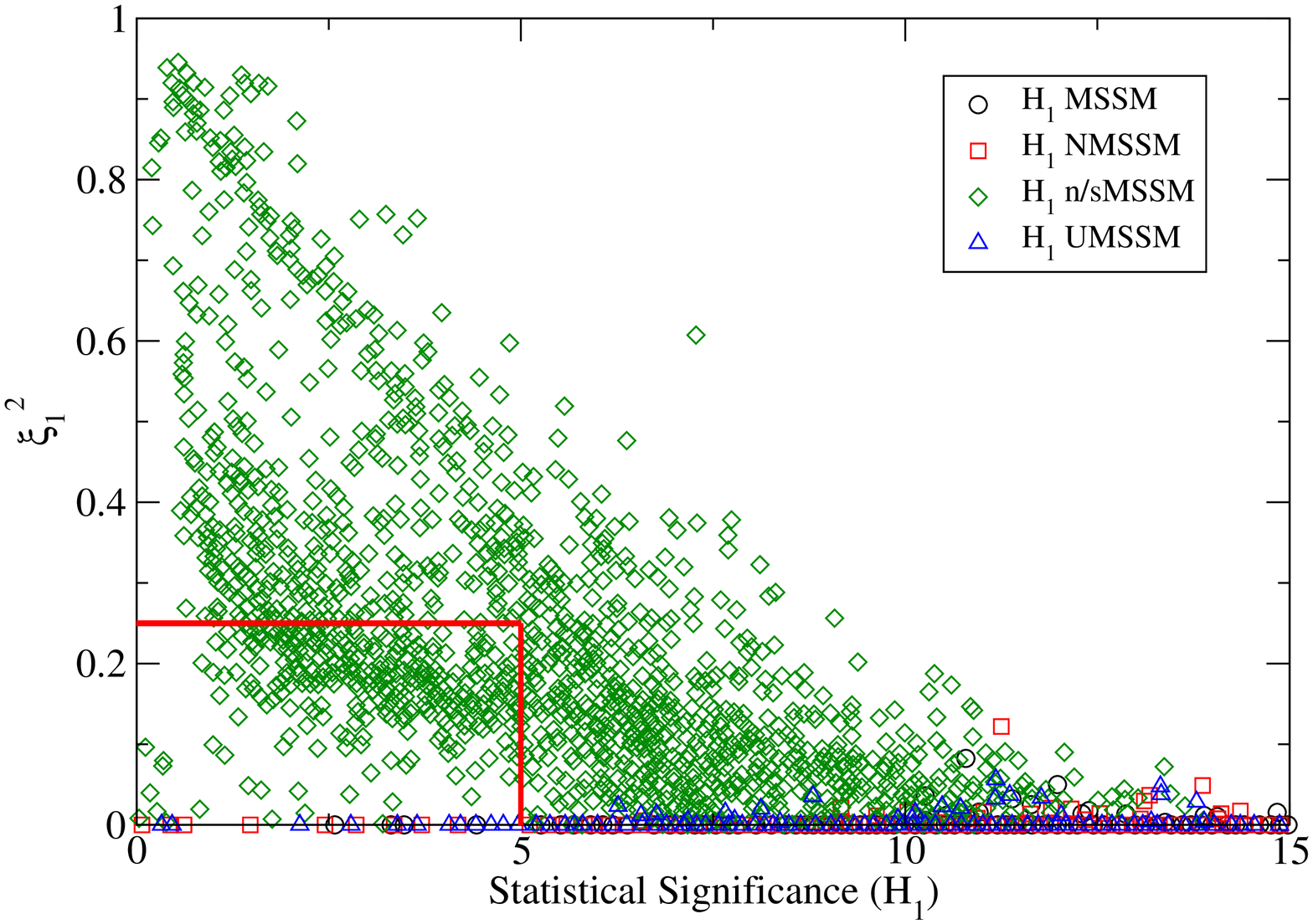}
\includegraphics[angle=-90,width=0.49\textwidth]{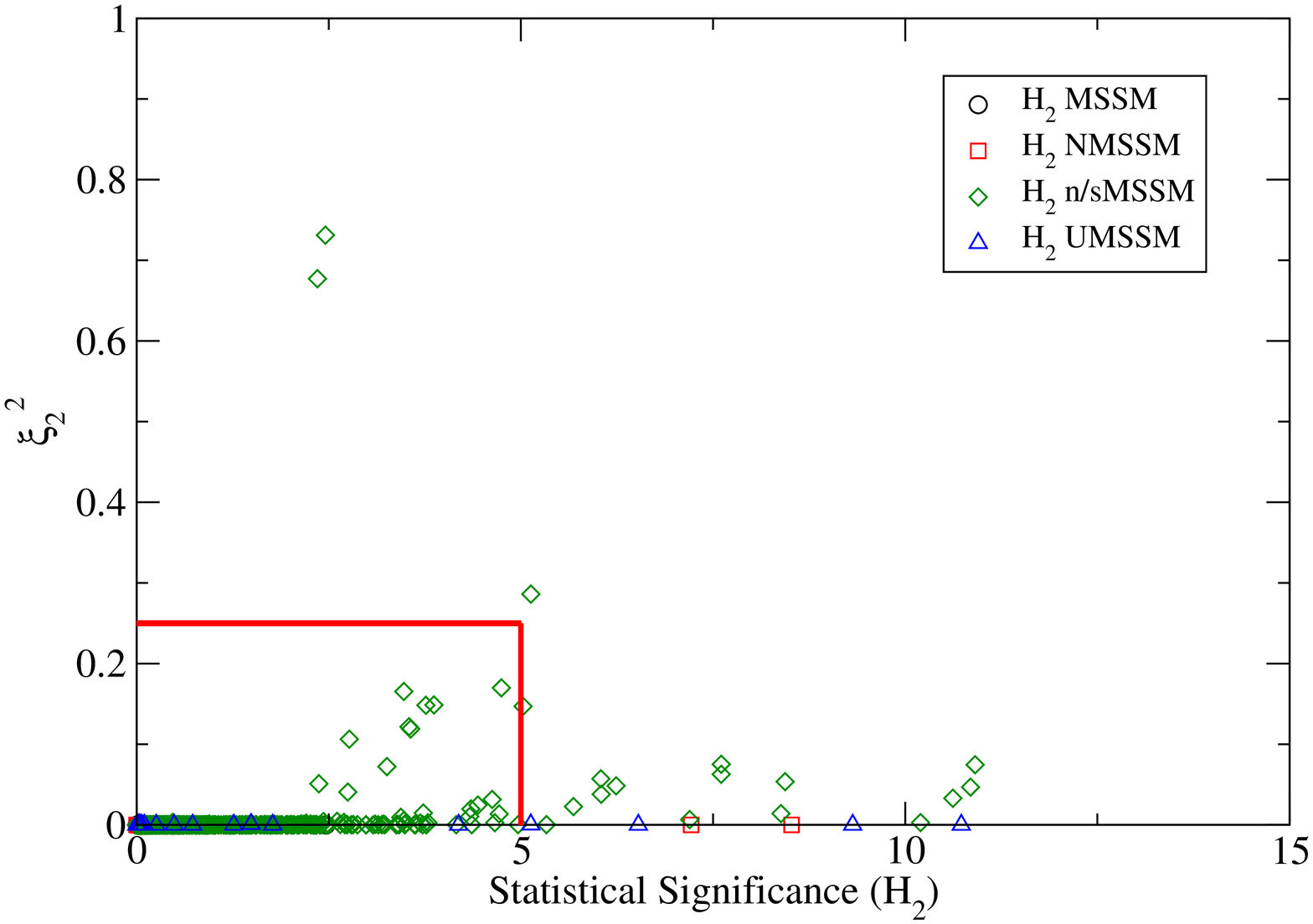}
(a)\hspace{0.48\textwidth}(b)
\caption{(a) Statistical signal significance of the expected Higgs signal at the LHC vs. the invisible Higgs decay fraction in WBF.   (b) For points where the lightest Higgs cannot be observed (i.e. the region inside the red box), we show the significance vs. invisible decay fraction for the second lightest Higgs.  Cases where $H_2$ can be discovered while $H_1$ remains elusive are possible; here, the lightest Higgs is typically light and dominantly singlet, making $H_2$ the MSSM-like lightest Higgs boson.}
\label{fig:sig-inv}
\end{center}
\end{figure}

The Higgs boson may be detected only by these indirect means in some scenarios (i.e. most cases of the n/sMSSM) while direct observation through traditional channels are favored in other scenarios such as the MSSM or UMSSM.  However, it is possible that both direct and indirect methods fail to find the Higgs boson.  Of the 2000 points generated randomly in each model, we find that in the n/sMSSM, neither Higgs boson is discoverable with the given luminosities with either method in 457 instances \footnote{Here, we require $S/\sqrt{B} < 5$ and $\xi_i^2 < 0.25$ for non-discovery.  In principle, points where both the statistical significance and invisible fraction are close to the expected limits may be discoverable by combining the statistics of the two methods.  The `no-lose theorem' for Higgs discovery may be spoiled by considering only these criteria.  However, it may be possible to observe a signal from $H\to A_1 A_1$ with $A_1 A_1\to b \bar b b \bar b$ or $A_1 A_1\to b \bar b \tau^+ \tau^-$ \cite{ref:han-huang}, although this is difficult for $m_{A_1} < 2 m_{b}$ \cite{Moretti:2006hq}.}.  We illustrate this point in Fig. \ref{fig:sig-inv}, where we show the statistical significance and the $\xi_i^2$ parameter for the lightest Higgs boson.  For the MSSM, NMSSM and UMSSM there were 5, 12 and 14 instances, respectively, for which the Higgs bosons are undiscoverable with these criteria with $100\text{ fb}^{-1}$ of integrated luminosity.  The possibility of discovering both light Higgs bosons in these models is low; we found only two instances in the NMSSM, five in the n/sMSSM and none in the MSSM and UMSSM in our scan.  However, it may be remotely possible to discover the second lightest Higgs boson in the n/sMSSM without discovering $H_1$, as we found 18 such points (only 2 such points were found for the NMSSM and 5 for the UMSSM).  This is a consequence of $H_1$ being very light and dominantly singlet while $H_2$ appears as the MSSM-like lightest Higgs.

\begin{table}[htdp]
\caption{Statistical signal significance and relative invisible Higgs branching fractions in WBF, $\xi_i^2$, in the MSSM and extended models in the scenarios of Fig. \ref{fig:illust}(a,b).}
\begin{center} 
\begin{tabular}{|c|c||ccc||ccc|}
\hline
	&	&	&	(a) &	&&(b)&\\
\hline
Model	&	MSSM&	N&	n/s &	U&	N&	n/s &	U\\
\hline
$m_{H_1}$& 127	&	127	&	72 	&	126	&	90&	131	 &114	\\
$m_{H_2}$& 1146	&	474	&	141 	&	187	&	133&209	 &141	\\
$m_{H_3}$& N/A	&	1354&	1030 &	1147&	1110&914	 &1147	\\
$m_{A_1}$& N/A	&	1089&	54 	&	N/A	&	134&197	 &N/A	\\
$m_{A_2}$&1146	&	1353&	 1032&	1147&	1110&908	 &1147	\\
$m_{\N_1}$&  85 	&	84	&	 34	&	84	&	88&	34	 &70	\\
\hline
$S_{H_1}/\sqrt{B}$&	10.2	&10.2&0.0	 &10.2		&2.4	&0.4	 &6.8\\
$S_{H_2}/\sqrt{B}$&	0.0	&0.0	&0.8	 	&0.2			&9.9	&1.2	 &6.1	\\
\hline
$\xi_{1}^2$&0.00	&0.00	&0.01	 &0.00	&0.00	&0.84	 &0.00	\\
$\xi_{2}^2$&0.00	&0.00	&0.86	 &0.00	&0.00	&0.00	 &0.00	\\
\hline
\end{tabular}
\end{center}
\label{tbl:illust}
\end{table}%

We further illustrate the two scenarios of Fig. \ref{fig:illust} in Table \ref{tbl:illust}, which gives the neutral Higgs and lightest neutralino masses along with the expected statistical significances of Higgs discoveries at the LHC.  The invisible Higgs modes are dominant in the n/sMSSM for this decoupled example and provide the only way to discover the Higgs in that scenario (the significances of visible modes are below one sigma).

Even if the methods for detecting an invisibly decaying Higgs can be extended to lower Higgs masses (below 100 GeV), they would not provide a good way to probe an invisible Higgs boson with mass below the LEP limit as the $VVH_i$ coupling is then required to be quite small to satisfy the LEP bound.  

The ILC can provide improved opportunities to discover singlet models \cite{Han:2004yd}.  An invisibly decaying Higgs boson can be detected at the ILC via the dominant production by $Z$-Higgstrahlung.  Since the incoming beam energies are known, the kinematics of the final state can be fully reconstructed by the observable $Z$ decay products \cite{ref:bargerzerwascheung}.  The recoil mass spectrum from the lepton pair can determine the mass of an invisible Higgs with an uncertainty of ${\cal O}(100 \text{ MeV})$ \cite{Abe:2001wn}.

\section{Coupling measurements}\label{sect:coupmeas}

\begin{figure}[htbp]
\begin{center}
\includegraphics[angle=-90,width=0.49\textwidth]{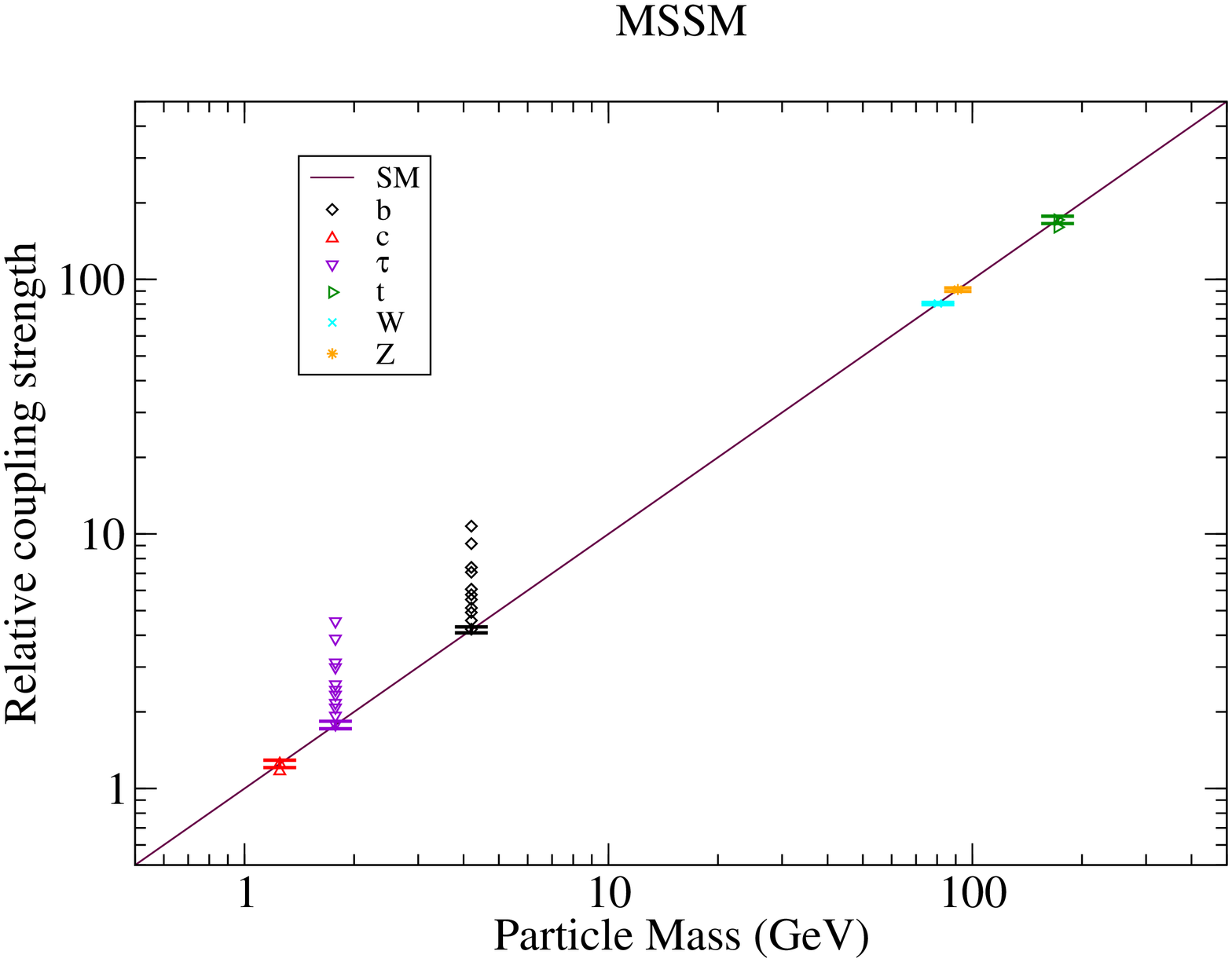}
\includegraphics[angle=-90,width=0.49\textwidth]{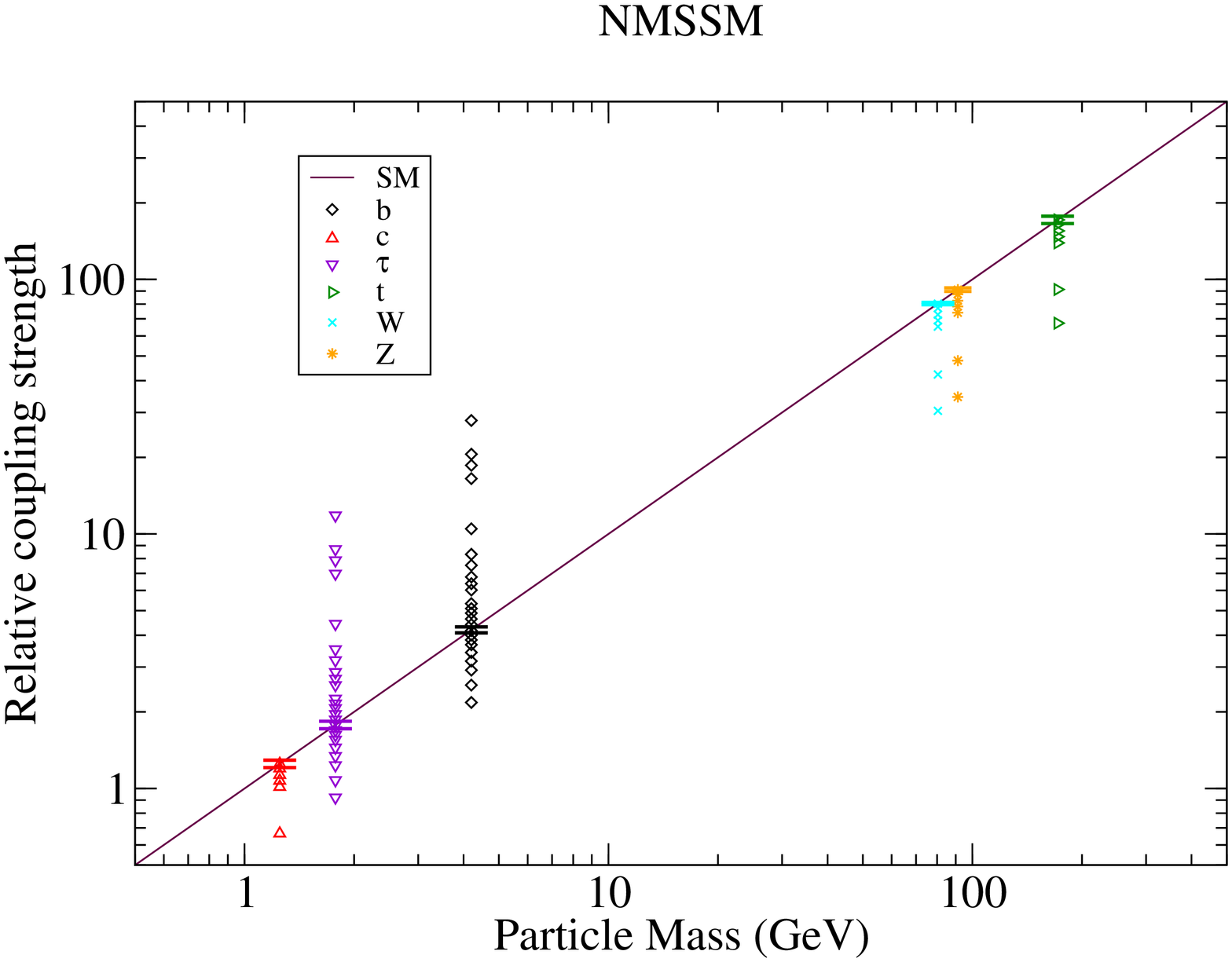}
(a)\hspace{0.48\textwidth}(b)
\includegraphics[angle=-90,width=0.49\textwidth]{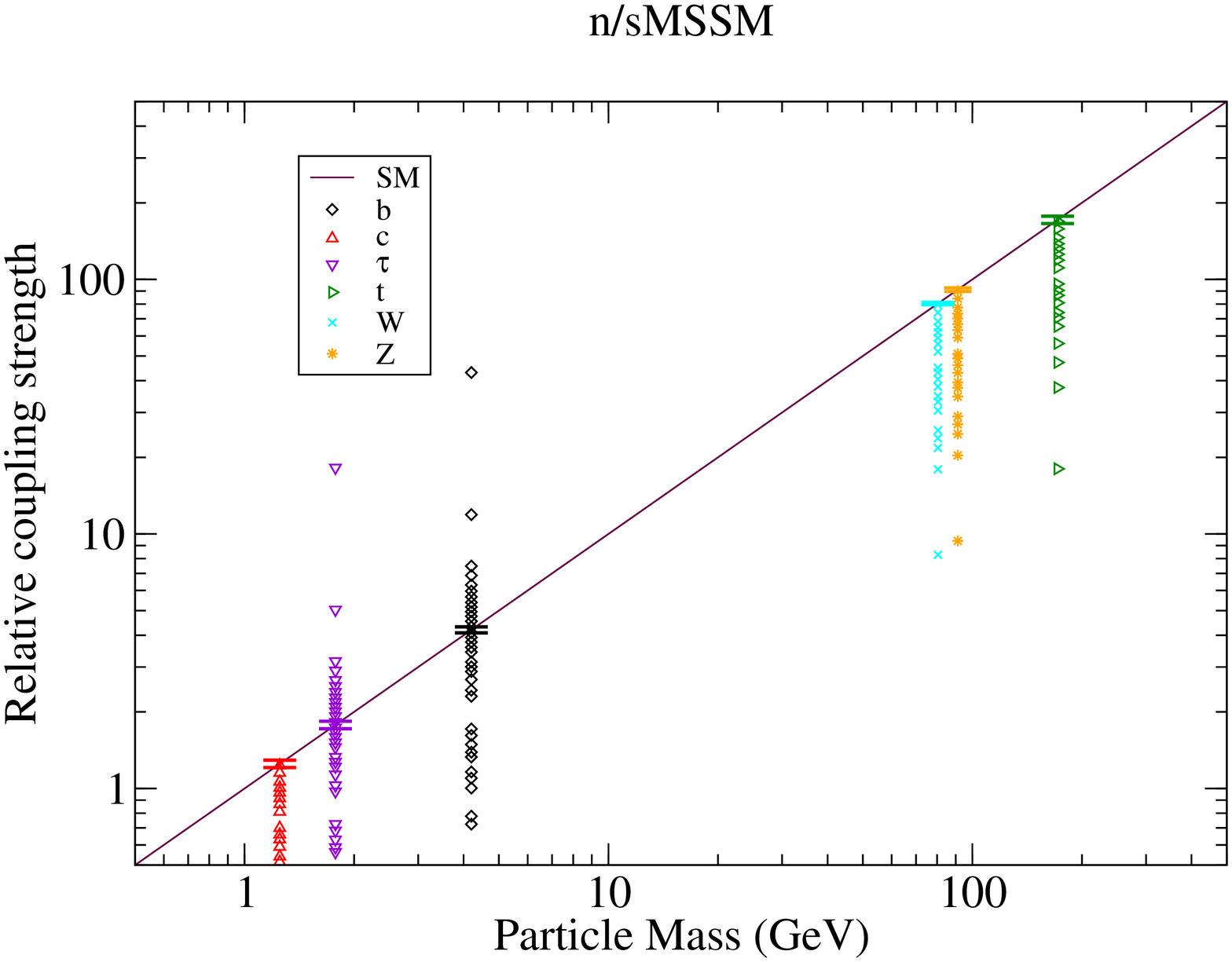}
\includegraphics[angle=-90,width=0.49\textwidth]{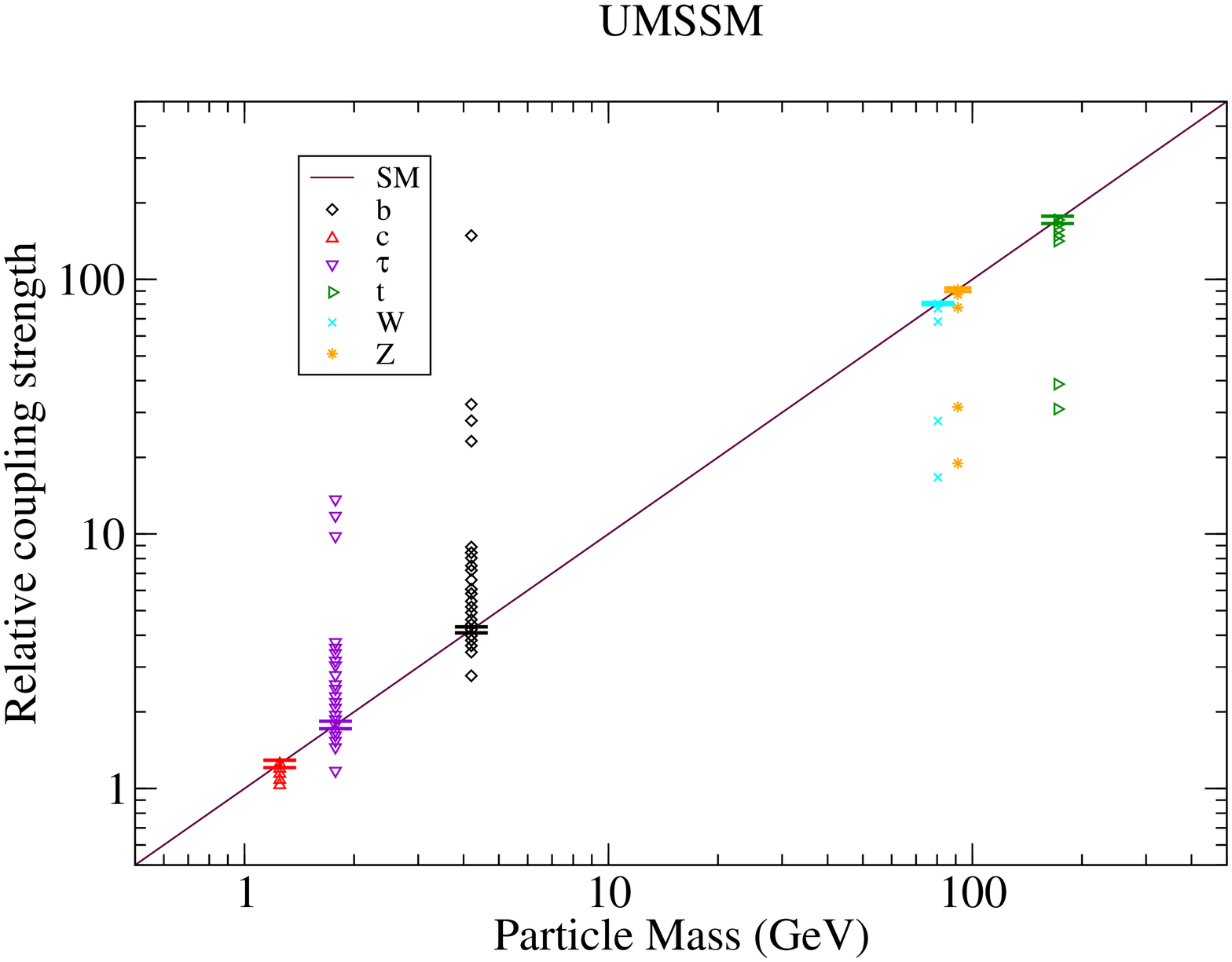}
(c)\hspace{0.48\textwidth}(d)
\caption{Comparison of the relative Higgs couplings in the MSSM and singlet models vs. the SM (straight line).  The expected uncertainties in coupling measurements at the ILC \cite{acfa-ilc} are also shown by the gaps between the horizontal bars centered on the SM values. }
\label{fig:coupcomp}
\end{center}
\end{figure}

Measurement of the gauge and fermion couplings of a Higgs boson can provide verification of the SM Higgs mechanism.  However, due to the extra singlet Higgs field, the couplings to Higgs bosons can be significantly different from SM predictions, making verification more difficult.  In Fig. \ref{fig:coupcomp}, we show the relative Higgs coupling dependence on the particle mass for Higgs bosons in the MSSM mass range $90\text{ GeV} \le m_{H_i}\le 135$ GeV; the SM values are given by the solid line and the expected coupling measurements uncertainties at the ILC \cite{acfa-ilc} are represented by the gaps between horizontal bars centered on the SM values.

In the MSSM, the couplings to gauge bosons and the charm and top quarks are similar to those of the SM, while the $\tan \beta$ enhancement for $\tau$-lepton and b-quark couplings is evident in Fig. \ref{fig:coupcomp}.  In the singlet extended models, the mixing can significantly change the couplings from the predictions of the MSSM.  In particular, due to singlet-doublet mixing the Higgs boson couplings to $W$, $Z$ and $t$ may be substantially reduced.  Deviation of the couplings from SM predictions may provide a smoking gun for the existence of a singlet.

In the MSSM the gauge couplings satisfy the relations
\bea
\xi^2_{VVH_1}&=&\xi^2_{VH_2A_2}=\sin^2(\alpha-\beta),\\
\xi^2_{VVH_2}&=&\xi^2_{VH_1A_2}=\cos^2(\alpha-\beta).
\label{eq:mssmcoup}
\eea
However, the extended singlet models do not obey these relations since the CP-odd rotation angle $\beta$ no longer diagonalizes the $3\times3$ CP-odd mass matrix.  The xMSSM couplings are given by
\bea
\xi^2_{VVH_i}&=&(R^{+}_{i1}\cos \beta+R^{+}_{i2}\sin \beta)^2,\\
 \xi^2_{VH_iA_j}&=&(R^{+}_{i1}R^{-}_{j1}-R^{+}_{i2}R^{-}_{j2})^2.
\eea

\begin{figure}[htbp]
\begin{center}
\includegraphics[angle=-90,width=0.49\textwidth]{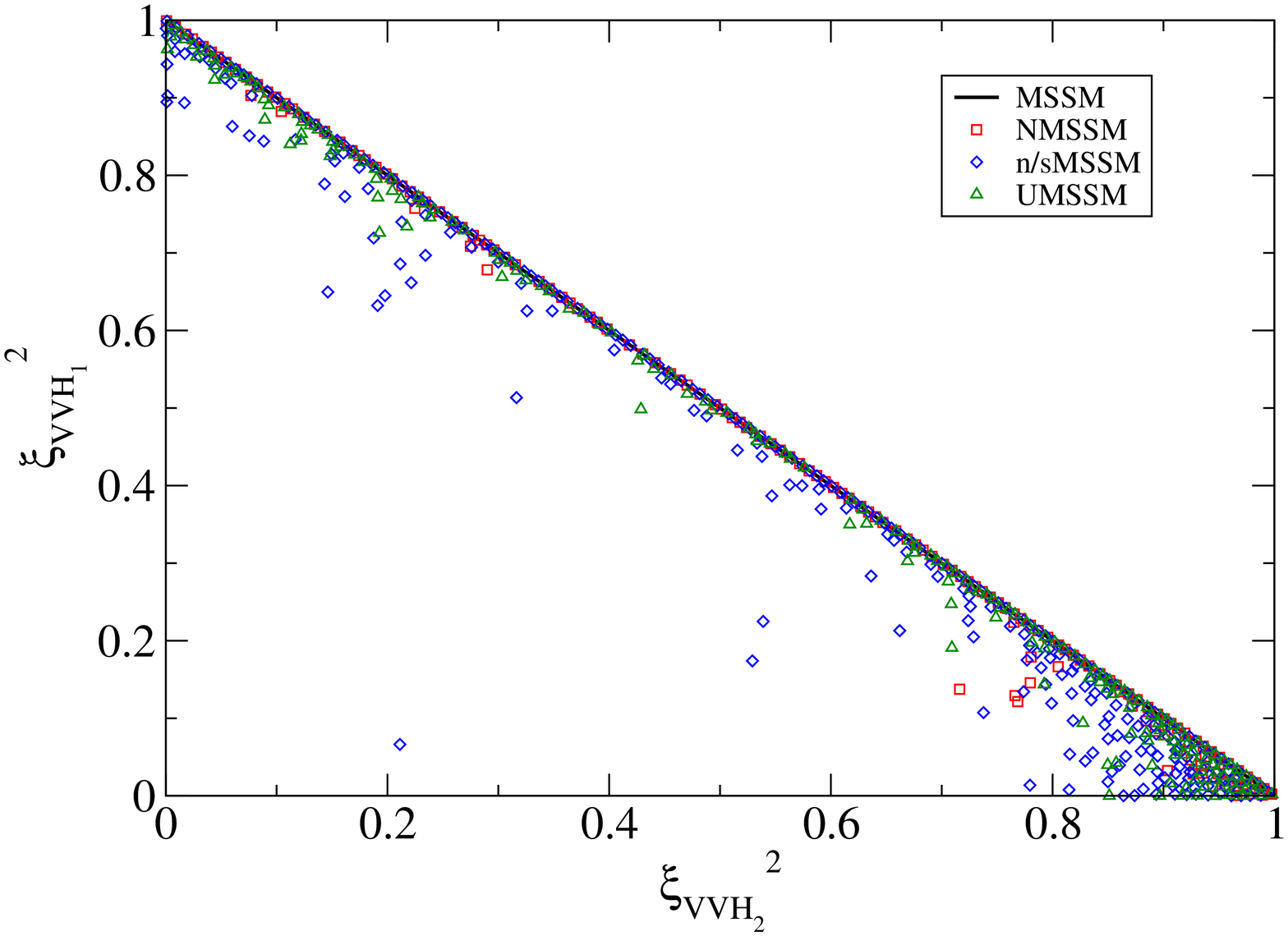}
\includegraphics[angle=-90,width=0.49\textwidth]{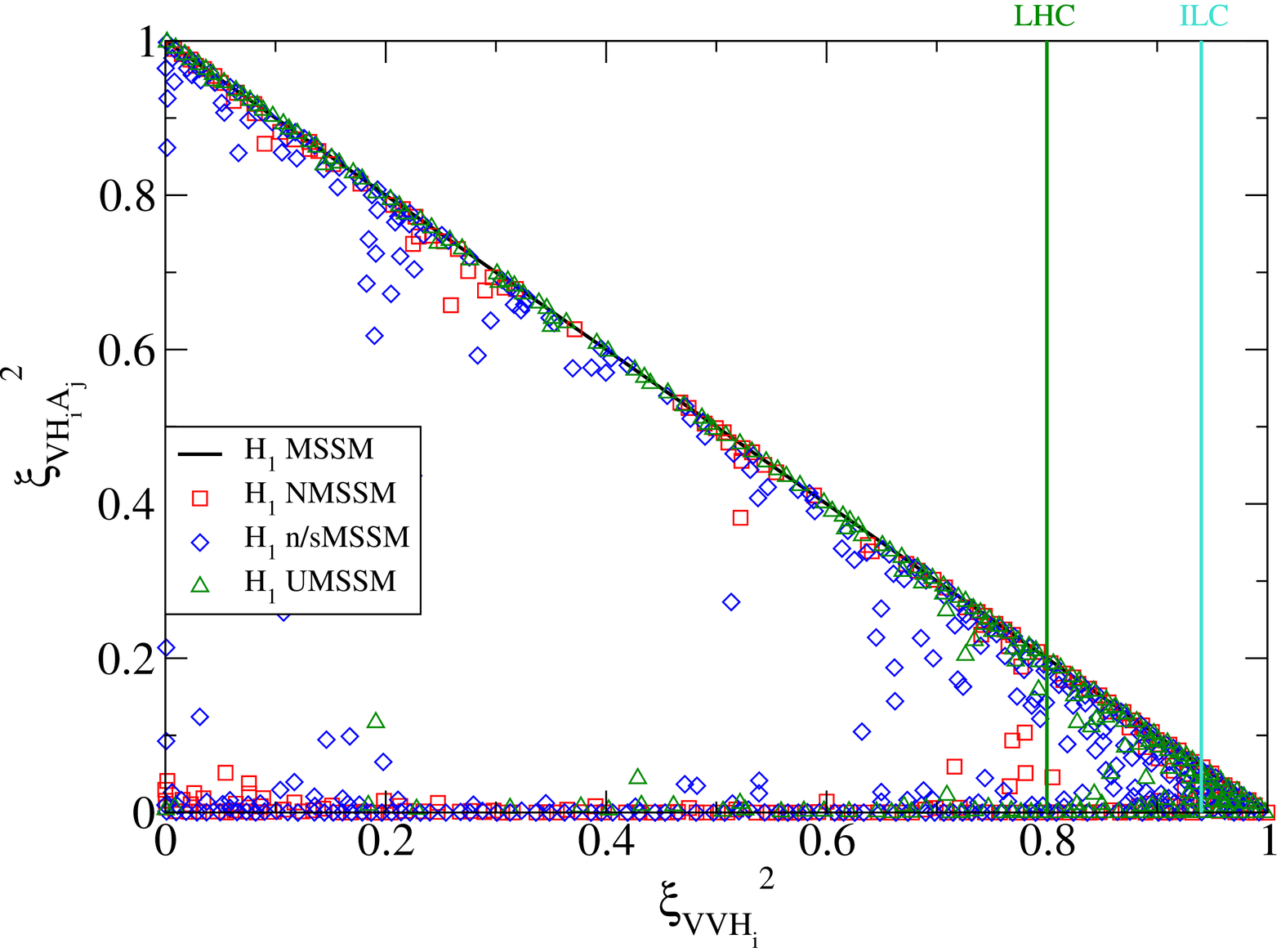}
(a)\hspace{0.48\textwidth}(b)
\caption{Comparison of the relative couplings of a Higgs boson with mass in the range $90\lesssim M_H\lesssim 135$ GeV.  The coupling complementarity between the $VVH_i$ and $V H_i A_j$ vertices does not hold in MSSM extensions due to the extra singlet (where $A_j$ is chosen to be the dominantly MSSM-like CP-odd Higgs boson).  The vertical lines in (b) represent the 1$\sigma$ uncertanties on $\xi_{VVH_i}^2$ expected from measurements at the LHC and ILC, with central values centered on unity for $\xi_{VVH_i}^2$.}
\label{fig:coup}
\end{center}
\end{figure}

We test the coupling equality of Eq. (\ref{eq:mssmcoup}) in the extended models and find that generally the $VVH_1$, $VVH_2$  couplings relative to the SM do not appreciably differ from the MSSM relations, as can be seen in Fig. \ref{fig:coup}a.  Over much MSSM parameter space (i.e. $A_{\lambda} > 200\text{ GeV}$), the $VVH_1$ relative coupling is very close to unity.  In Fig. \ref{fig:coup}a, we therefore denote the MSSM as a solid black line. Assuming a SM-like coupling, combined ATLAS and CMS data with 800 fb$^{-1}$ of combined integrated luminosity are expected to measure Higgs couplings to vector bosons to a 1$\sigma$ accuracy of order 10-30\% of $g_{\rm SM}$ in the mass range $110\text{ GeV} \le M_h \le 190\text{ GeV}$ \cite{Duhrssen:2004uu}.  The ILC with $\sqrt{s}=500$ GeV with 500 fb$^{-1}$ will provide more precise measurements of the couplings to ${\cal O}(2-5\%)$ for light Higgs bosons ($M_h \sim 120$ GeV) \cite{Abe:2001np}.  The expected experimental uncertainties are shown in Fig. \ref{fig:coup}b.  The deviations of the scaled squared $VVH_i$ couplings of xMSSM models from the MSSM expectations are not larger than the measurement uncertainties expected at the LHC or ILC.  

If a Higgs boson is discovered in the mass range of the lightest Higgs in the MSSM, $90 \lesssim M_H \lesssim 135 $ GeV, the equalities of Eq. (\ref{eq:mssmcoup}) involving the heaviest CP-odd state, $A_2$, may be considered as a way to differentiate the extended models from the MSSM.  In Fig. \ref{fig:coup}b, we present the scaled $VVH_i$ coupling-squared against the scaled $VH_iA_j$ coupling-squared (where $A_j$ is chosen to be the dominantly MSSM-like CP-odd Higgs boson) for the CP-even Higgs masses in this mass range.  We also plot the expected ILC and LHC uncertainties from unity for the scaled $VVH_i$ coupling in Fig. \ref{fig:coup}b, assuming central measured values of unity for $\xi_{VVH_1}^2$.  A majority of the points fall at  $\xi_{VH_i A_j}^2 = 0$ and $\xi_{VVH_i }^2 = 1$ where the MSSM Higgs sector is decoupled and mimics the SM \cite{Barger:1992zy}.  The high density of points in this region has been reduced for clarity in this figure.   The $V H_i A_j$ coupling is often suppressed below the MSSM line since CP-odd mixing can occur in addition to CP-even mixing. 

Distinguishing the UMSSM from the MSSM with coupling measurements will be challenging as the electroweak constraints on the $Z-Z'$ mixing angle $\alpha_{ZZ'}$ require the lightest CP-even Higgs to be similar to that of the MSSM \cite{Barger:2006dh}.  Therefore, the $Z H_{i} A_j$ coupling is similar to that of the MSSM when $H_{i}$ is the heavy Higgs in the MSSM, and the differences are very small and below the expected sensitivity of the LHC, ILC combined analysis.  However, other avenues of discovery are possible in this model such as the $Z'$ boson and neutralino cascades \cite{Barger:2006kt}.

\section{Conclusions}\label{sect:concl}

Higgs singlet extensions of the MSSM provide well motivated solutions to the hierarchy problem.  The Higgs singlet increases the number of CP-even and CP-odd Higgs states which lead to interesting collider phenomenology.  Specifically, we find the following in the extended models:

\begin{itemize}

\item The lightest Higgs can be lighter than the LEP limit of 114 GeV with reduced Higgs couplings to SM fields; the production rates of these states in visible channels are often below the rates for the lightest MSSM Higgs boson.

\item In most of the parameter space of the extended models, at least one Higgs state is discoverable via traditional SM decay modes at the $5\sigma$ level with $>30\text{ fb}^{-1}$ luminosity at the LHC.  Direct observation of the lightest Higgs is favored for the MSSM, NMSSM and UMSSM.  In the n/sMSSM the traditional discovery modes can be spoiled by the decays to invisible states such as neutralinos.

\item Indirect Higgs observation can be employed for the n/sMSSM where invisible Higgs decays to neutralino pairs are often dominant.  Cascade decays of $H_1\to \N_2 \N_1 \to2 \N_1 + \nu \bar \nu$ and $H_2\to 2 H_1 \to 4 \N_1$ can contribute to the invisible decay modes.

\item A Higgs boson in the mass range 114-135 GeV can be discovered directly or indirectly in most cases.  In some cases, the second Higgs state can be discovered while the lightest cannot.  However, in the n/sMSSM, a substantial portion, ${\cal O}(20\%)$, of the parameter space does not allow discovery of a CP-even Higgs state, posing a difficult scenario for Higgs physics at the LHC.  

\item The Yukawa coupling to fermions can be either $\tan \beta$ enhanced as in the MSSM or suppressed or enhanced in xMSSM models due to the more complicated mixing patterns involving singlets.  

\item Measurements of the $VVH_i$ and $V H_i A_j$ couplings can in principle differentiate the singlet models from the MSSM.  In the MSSM, these couplings are equivalent due to the simple $2\times 2$ Higgs rotation matrices.  However, the equivalence does not hold in the singlet models since the rotation matrix is more complicated due to singlet mixing.

\item Scenarios exist where the singlet extended models are difficult to differentiate from the MSSM using only the Higgs sector.  However, complementary avenues are available through the discovery of a $Z'$ boson in the UMSSM or extended neutralino cascade decays due to the different neutralino spectra in singlet extended models.   

\end{itemize}

\begin{acknowledgments}
This work was supported in part by the U.S.~Department of Energy under grant No. DE-FG02-95ER40896, by the Wisconsin Alumni Research Foundation and by the Friends of the IAS.  VB thanks the Aspen Center for Physics for hospitality during this work.  We thank T. Han and G. Huang for helpful discussions and J. Hewett for information.
\end{acknowledgments}

\end{document}